\documentclass[preprint,3p]{elsarticle}
\usepackage{amsmath}
\usepackage{comment}
\usepackage{color}
\usepackage{braket}
\usepackage{amssymb}
\usepackage{makeidx}

\def\Tr{\mathrm{Tr}}
\def\diag{\mathrm{diag}}

\def\dd{\mathrm{d}}

\def\simge{\mathrel{%
    \rlap{\raise 0.511ex \hbox{$>$}}{\lower 0.511ex \hbox{$\sim$}}}}
\def\simle{\mathrel{
    \rlap{\raise 0.511ex \hbox{$<$}}{\lower 0.511ex \hbox{$\sim$}}}}

\newcommand \be{\begin{eqnarray}}
\newcommand \ee{\end{eqnarray}}
\newcommand{\del}{\partial}

\def\XXint#1#2#3{{\setbox0=\hbox{$#1{#2#3}{\int}$}
\vcenter{\hbox{$#2#3$}}\kern-.5\wd0}}

\begin{document}

\begin{frontmatter}




\title{Unveiling the significance of eigenvectors in diffusing non-hermitian matrices by identifying the underlying Burgers dynamics}

\author[a1]{Zdzislaw Burda}
\ead{zdzislaw.burda@agh.edu.pl}
\author[a2]{Jacek Grela\corref{cor1}}
\ead{jacekgrela@gmail.com}
\author[a2]{Maciej A. Nowak}
\ead{nowak@th.if.uj.edu.pl}
\author[a2]{Wojciech Tarnowski}
\ead{wojciech.tarnowski@uj.edu.pl}
\author[a2]{Piotr Warcho\l{}}
\ead{piotr.warchol@uj.edu.pl}

\address[a1]{AGH University of Science and Technology, Faculty of Physics 
and Applied Computer Science, al. Mickiewicza 30, PL--30059 Krak\'ow, Poland}
\address[a2]{M. Smoluchowski Institute of Physics and Mark Kac Complex Systems Research Centre, Jagiellonian University,  PL--30348 Krak\'ow, Poland}

\cortext[cor1]{Corresponding author.}
\author{}

\address{}

\begin{abstract}

Following our recent letter~\cite{BGNTW1}, we study in detail an entry-wise diffusion of non-hermitian complex matrices. We obtain an exact partial differential equation (valid for any matrix size $N$ and arbitrary initial conditions) for evolution of the averaged extended characteristic polynomial. The logarithm of this polynomial has an
interpretation of a potential which generates a Burgers dynamics in quaternionic space.
The dynamics of the ensemble in the large $N$ is completely determined by the coevolution of the spectral density and a certain eigenvector correlation function. This coevolution is best visible in an electrostatic potential of a quaternionic argument built of two
complex variables, the first of which governs standard spectral properties while the second unravels the hidden dynamics of eigenvector correlation function.   
We obtain general large $N$ formulas for both spectral density and 1-point eigenvector correlation function valid for any initial conditions. We exemplify our studies by solving three examples, and we verify the analytic form of our solutions with numerical simulations. 

\end{abstract}

\begin{keyword}
Non-hermitian random matrix model \sep Diffusion equation \sep Burgers equation \sep Characteristic polynomial

\PACS{05.10.-a,
02.10.Yn,
02.50.Ey,
47.40.Nm}
\end{keyword}

\end{frontmatter}

\section{Introduction}
The concept of matrices filled with entries subject to the diffusion process was first introduced by Dyson~\cite{DYSON} and applied in the context of both Gaussian Unitary Ensemble (GUE) and Circular Unitary Ensemble (CUE). The arising Coulomb gas analogy had a major impact on understanding of random matrices~\cite{FORRESTERBOOK}. Today the marriage of stochastic processes and random matrices still brings several new insights. Examples include the study of determinantal processes~\cite{MAJUMDAR,KATORI, SCHEHR}, Loewner diffusion~\cite{LOEWNER} or the fluctuations of non-intersecting interfaces in thermal equilibrium~\cite{NADAL}.

Recently, several authors~\cite{BN1,BGNW1} have approached the diffusion in the GUE from a new perspective. They found a viscid complex Burgers equation for the logarithmic derivative of the averaged characteristic polynomial $f_{N}(z, \tau)\equiv \frac{1}{N} \partial_z \ln U_{N}(z,\tau)$ (associated with a Hermitian matrix filled with entries performing Brownian motion in the complex space):
\begin{align}\label{eq:bev}
	\partial_\tau f_N (z,\tau) + f_N(z,\tau) \partial_z f_N(z,\tau) = -\frac{1}{2N} \partial^2_{z} f_N(z,\tau),
\end{align}
where $\tau$ is the diffusion time and $z$ is a complex variable. The role of viscosity is played by the inverse size of the matrix $N$.
In the $N\to \infty$ limit, $f_{N}(z, \tau)$ becomes the Green's function $G(z,\tau)$ and the partial differential equation becomes inviscid:
\begin{align}
	\partial_\tau G(z,\tau) + G(z,\tau) \partial_z G(z,\tau) = 0.
	\label{BurgersGUE}
\end{align}
A solution of the latter equation (by  the method of characteristics) requires an introduction of shocks, which turn out to coincide with the edges of the spectra. 
This phenomenon leads to a novel interpretation of known matrix results, since   microscopic universal behavior of the spectra emerges as an expansion around the shock wave of the viscous equation. Nontrivial initial conditions give rise to shock collisions which are equivalent to the merging of the spectrum boundaries. This way, not only Airy but also the Pearcey functions are captured in the same formalism. A similar Burgers equation was also obtained for the Wishart ensemble and chiral GUE yielding a universal scaling associated with the Bessoid function~\cite{BNWWISHART2}. The equivalent  phenomenon appears also at the level of CUE diffusion, providing new insight for order-disorder transition of Wilson loops in Yang-Mills theory~\cite{NEUBERGER}. 

Recently, this program got extended to the realm of matrices with complex eigenvalues~\cite{BGNTW1}. The success of such an extension is a priori surprising, since hermitian and non-hermitian matrix models seem to be hardly comparable. 
In the former, the hermiticity condition confines the eigenvalues to the real axis. In the latter, there are no constraints, and the eigenvalues spread over the whole complex plane. Furthermore, non-hermitian models develop discontinuities at the spectral density boundaries, a feature observed even for the well-known Ginibre Ensemble~\cite{GG}, for which the spectral density is given by:
\begin{align}
	\rho(z,\bar{z}) = \frac{1}{\pi} \theta(1-|z|),
\end{align} 
whereas e.g. in the GUE case, the Wigner semicircle is continuous across the spectral edge, and only the derivatives are discontinuous. Other differences arise when one considers the evolution of the diffusing matrices. In the hermitian case, the evolution is determined by the initial eigenvalues only, whereas in the non-hermitian models, the information on initial eigenvectors  additionally affects the shape of the spectral density.

This paper is a continuation and an extension of the ideas of non-hermitian diffusion announced briefly in~\cite{BGNTW1}.

Our approach is rooted in the standard electrostatic analogy, but requires a novel setting (we call it the quaternionic method), which can be viewed as an extension of the standard Dysonian strategy applied originally to the hermitian (or unitary) matrix diffusion case. 
The main objects of our interest are  the spectral density (obtained from the resolvent) {\it and} a certain one-point eigenvector correlation function.
We stress that the aforementioned eigenvector correlator is crucial for understanding the diffusion process of non-hermitian matrices. We also point out, why the importance of this correlator was disregarded in majority of the studies of non-hermitian random matrix models.  

The basic object of our studies is an averaged ``extended'' characteristic polynomial (AECP). An extension follows from an introduction of two pairs of complex variables (compared to one complex variable  in standard treatments). Surprisingly, AECP obeys a certain partial differential equation akin to the diffusion equation, for arbitrary size of the matrix and for arbitrary initial conditions, and is exactly integrable. The diffusion happens in the auxiliary plane ``perpendicular'' to the complex plane where the eigenvalues reside. In the large matrix size limit, the logarithm of the AECP can
be viewed as an electrostatic potential  and its derivatives with respect to the two complex variables yield a pair of coupled Burgers-like equations for the non-hermitian Green's function and eigenvector correlation function. 
We would like to mention, that in the standard electrostatic analogy~\cite{STERN,FIODOROV,BROWN} the "second variable" is treated as an infinitesimal regulator only. 
This is the reason why the dynamics, as a function of this variable, remained hidden, and the complementary information on the eigenvector correlator co-evolving with the spectra  was absent. 

To illustrate our findings, we consider a couple of examples of initial conditions. In most of them we demonstrate the explicit solutions of Burgers equations, obtaining formulas for the spectral density,  eigenvector correlators and the electrostatic potential. We note that an inspection of the Burgers-like equation identifies the shock line with the non-holomorphic sector of the spectral density. Moreover, we show the insensitivity of the shock formation to the initial condition chosen. This hints to a lack of truly new universality classes in this type of models, which is furthermore corroborated in the study of the universal behavior in the vicinity of the spectral collision in one of the examples. 

The paper is organized as follows. In Section \ref{mdiff} we define Dysonian non-hermitian diffusion. In Section \ref{quatmeth} we briefly review the electrostatic analogy and the quaternionic method. We proceed in Section \ref{aecp} by deriving the partial differential equation for the AECP and presenting its integral representation. In Section \ref{pairburg} we derive a pair of coupled Burgers equations for the on- and off-diagonal parts of the quaternionic Green's function (in the large $N$ limit) thus making a link to the quaternionic method. Subsequently we solve them with the method of complex characteristics. Finally, we obtain an implicit solution to the equation for the potential in terms of the Hopf-Lax formula and derive large $N$ formulas for the spectral density, eigenvector correlation functions and the boundary of the spectrum valid for an arbitrary initial matrix. Section \ref{examples} is devoted to the examples of a) Ginibre, b) spiric and c) 1-band non-normality matrices. We apply previously described methods to these cases, depict the characteristics picture and obtain the large $N$ limit spectral density and eigenvector correlators. We also comment on critical behavior of the EACP. Section \ref{stochasticeq} is devoted to a curious observation by Osada~\cite{OSADA1}, which actually has triggered our interest in the diffusion of the Ginibre ensemble. We provide also an explanation of the Osada's observation. Section \ref{conclusions} summarizes the paper and outlines some possibilities of further investigations of the observed patterns. 

Three appendices hide technicalities: Appendix A demonstrates the derivation of the key diffusion equation for an AECP, Appendix B clarifies the link between Ginibre and Wishart/chiral ensembles, and Appendix C determines weights and normalizations needed for establishing the universal scaling at the shock. 

\section{Dysonian non-hermitian diffusion}
\label{mdiff}

Consider a non-hermitian $N\times N$ matrix $X = (X_{ij})_{i=1,\ldots,N \; j=1,\ldots,N}$
whose elements $X_{ij}=x_{ij}+iy_{ij}$ undergo
$2N^2$-dimensional Brownian motion 
\begin{align}
\label{gp}
dx_{ij}(\tau) = \frac{1}{\sqrt{2N}} dB_{ij}^x(\tau), 
\quad dy_{ij}(\tau) = \frac{1}{\sqrt{2N}} dB_{ij}^y(\tau),
\end{align}
where $B_{ij}^{x}$ and $B_{kl}^y$ are independent Wiener processes. 
We restrict ourselves to deterministic initial conditions, that is we assume that each element of the matrix has a given initial value $x_{ij}=(x_{0})_{ij}$ and $y_{ij}=(y_{0})_{ij}$ for $\tau=0$. This condition can be concisely written as $X=X_0$, for $\tau=0$. Clearly, this model is a straightforward extension of the Dyson random walk \cite{DYSON} to the realm of non-hermitian random matrices.
 
The joint probability density for matrix elements evolves according to the 
$2N^2$-dimensional diffusion equation
\be
\del_\tau P(X,\tau) = \frac{1}{4N} \sum_{ij} ( \del^2_{x_{ij}} + \del^2_{y_{ij}}) P(X,\tau), \label{prob}
\ee 
with the initial condition $P(X,0) = \delta(X-X_0)$. 
The probability measure for random matrices at time $\tau$ is defined by
$d \mu_\tau(X) = \mathcal{D}[X] P(X,\tau)$ where 
$\mathcal{D}[X] = \prod_{ij} d x_{ij} d y_{ij}$, and the 
statistical averages by
\begin{align}
\langle F(X) \rangle_\tau = \int \mathcal{D}[X] P(X,\tau) F(X).
\end{align}
The hermitian version of the model, discussed by Dyson, reduces to a model
of evolution of eigenvalues. In that case eigenvectors can be integrated out.
What makes the non-hermitian extension interesting is that in addition
to eigenvalues one has to control also the evolution of eigenvectors. We present a systematic method to do so.

\section{Electrostatic analogy and quaternions}
\label{quatmeth}

In this section we briefly recall the method to calculate eigenvalue
distribution of random matrices in the limit $N\rightarrow \infty$.
The method is based on ``electrostatic'' analogy~\cite{STERN,FIODOROV,BROWN}. One defines a quantity
\begin{align}
\label{pot}
	\Phi(z,w,\tau) = 
	\frac{1}{N} \left < {\rm Tr} \log \left ( (z-X)(\bar{z} - X^\dagger) + |w|^2 \right ) \right >_\tau ,
\end{align}
which can be interpreted in the limit $w \rightarrow 0$ as an electrostatic 
potential of a cloud of $N$ electric charges interacting on the $z$-complex plane. 
The corresponding electric field is
\begin{align}
\label{GPhi}
	G\left(z, w,\tau\right) = \partial_{z} \Phi(z,w,\tau) = \frac{1}{N}
	 \left < {\rm Tr} \frac{\bar{z}-X^{\dagger}}{(z-X)(\bar{z}-X^{\dagger} )+|w|^{2} } \right >_\tau.
\end{align}
Identifying the real and imaginary part of as vector components
$G= (E_x - i E_y)/2$ one can rewrite the last equation in the vector notation as $\vec{E} = (E_x,E_y) = \vec{\nabla}_z \Phi$. The minus sign in front of $E_y$ and the scale factor $1/2$ in the relation of $G$ to electric field $\vec{E}$ is a matter of convention. 

We are interested in the eigenvalue distribution 
\be
\rho (z,\tau) \equiv \frac{1}{N}\left < \sum_i {\delta^{(2)} 
\left(z - z_{i}\right)}\right >_\tau,
\ee
where $z_i$'s are the eigenvalues of $X$. The limiting eigenvalue density can be calculated from the Gauss law 
\begin{align}
\label{gausslaw}
\rho (z,\tau) = \frac{1}{\pi} \partial_{\bar{z}} G\left(z,w,\tau\right)
\qquad , ~w\rightarrow 0 .
\end{align}
This relation follows from a standard representation of the complex Dirac delta function $\pi\delta^{(2)}\left(z-z_{i}\right)=\underset{|w| \rightarrow 0}{\lim}\frac{|w|^{2}}{\left(|w|^{2}+\left|z-z_{i}\right|^{2} \right)^{2}}$.
The expression in the brackets on the r.h.s. of (\ref{GPhi}) can be cast 
into the standard form of resolvent $(z-X)^{-1}$ 
at  the price of introducing $2N\times 2N$ matrices
\begin{align}
Q = \left ( \begin{matrix} z & -\bar{w} \\
	 w & \bar{z}  
	 \end{matrix} \right ) \ , \quad  
	 \mathcal{X} = \left ( \begin{matrix} X & 0 \\
	 0 & X^\dagger  
	 \end{matrix} \right ) ,
\end{align}
in place of the original $N\times N$ ones. The resolvent is a $2\times 2$
matrix 
\begin{align}
\label{quatgf}
\mathcal{G}(z,w,\tau) \equiv
\left( \begin{matrix} \mathcal{G}_{11} & \mathcal{G}_{1\bar{1}} \\
\mathcal{G}_{\bar{1}1} & \mathcal{G}_{\bar{1}\bar{1}}  
\end{matrix} \right ) =
\frac{1}{N} \left < \rm bTr \frac{1}{Q-\mathcal{X}} \right >_\tau ,
\end{align}
where the block-trace is defined as
$$
	\rm bTr \left ( \begin{matrix} A & B \\
	 C & D  
	 \end{matrix}\right ) = \left ( \begin{matrix} \Tr A & \Tr B \\
	 \Tr C & \Tr D  
	 \end{matrix}\right ).
$$
We refer to $\mathcal{G}(z,w,\tau)$ (\ref{quatgf}) as to
generalized Green's function or quaternionic resolvent 
\cite{JANIKNOWAK,JAROSZNOWAK}. Note, that we use the representation of the quaternion in terms of Pauli matrices, i.e. $Q=q_0 1_2 +i\sum_{j=1}^3 \sigma_j q_j$, so $z=q_0+iq_1$ and $-\bar{w}=q_2+iq_3$. The diagonal element of the quaternionic resolvent $\mathcal{G}_{11}$ is equal to $G(z,w,\tau)$ (\ref{GPhi}). The extended Green's function $\mathcal{G}(z,w)$ is an advantageous object since one can apply geometric series expansion
to $(Q-\mathcal{X})^{-1}$ which has a diagrammatic interpretation. This leads to a closed set of Dyson-Schwinger equations enumerating underlying planar Feynman diagrams. From these equations one can derive an exact form of the  Green's function in the limit $N\rightarrow \infty$, as well as matrix-valued addition and multiplication laws~\cite{BJN}. We mention that the quaternionic extension is equivalent to another approach known under the name of hermitization method \cite{GIRKO,FEINBERGZEE,CHALKERWANG}, in which the diagonal and off-diagonal blocks of matrices $Q$ and $\cal{X}$ are flipped before the block-trace operation. 

Having determined the quaternionic  resolvent $\mathcal{G}(z,w,\tau)$ 
one can determine the potential $\Phi(z,w,\tau)$ or vice versa, since
the two objects are related by a simple relation: 
\begin{align}
\label{G_dPhi}
 \mathcal{G} = \left ( 
 \begin{matrix} \partial_z \Phi & \partial_w \Phi \\
 -\partial_{\bar{w}} \Phi & \partial_{\bar{z}} \Phi
 \end{matrix} \right ).
\end{align}
As follows from (\ref{gausslaw}), the eigenvalue density can be derived from
the potential $\Phi(z,w,\tau)$ using the Poisson equation
\be
\label{Poisson}
\rho(z,\tau) = \frac{1}{\pi} \partial_{\bar{z}z} \Phi\left(z,0,\tau\right). 
\ee
It turns out that the potential $\Phi(z,w,\tau)$
encodes also information about the correlations of eigenvectors \cite{CHALKERMEHLIG}, being the special case of the Bell-Steinberger matrix \cite{BELL,SAVINSOK,FYODSAV}.
One defines the correlation function as\footnote{Note that we introduced an
additional $1/N$ factor as compared to the definition given in~\cite{CHALKERMEHLIG}
to obtain a limiting density.}
\begin{align}
\label{evs}
O(z,\tau) \equiv \frac{1}{N^2} \left < \sum_\alpha O_{\alpha\alpha} 
\delta^2(z-z_\alpha) \right >_\tau ,
\end{align}
with $O_{\alpha\beta} = \braket{L_\alpha | L_\beta}\braket{R_\alpha | R_\beta}$ where $\ket{L_\alpha}$ ($\ket{R_\alpha}$) are the left (right) eigenvectors of matrix $X$.
It can be shown~\cite{NOWAKNOER} that, in the $N \to \infty$ limit, this correlation function is related to the off-diagonal elements of the resolvent as
$O(z,\tau) = 
-\frac{1}{\pi} \mathcal{G}_{1\bar{1}}(z,0,\tau) \mathcal{G}_{\bar{1}1}(z,0,\tau)$.
Applying (\ref{G_dPhi}) we have
\be
\label{OV}
O(z,\tau) = \frac{1}{\pi} \left.
\partial_w \Phi(z,w,\tau) \partial_{\bar{w}} \Phi(z,w,\tau)\right|_{w=0} =
\frac{1}{\pi}  |V(z,0,\tau)|^2,
\ee
where $V(z,w,\tau) = \partial_w \Phi(z,w,\tau)$ is the velocity field,
which plays the same role in the $w$-complex plane as the electric field
$G(z,w,\tau) = \partial_z \Phi(z,w,\tau)$ in the $z$-complex plane. It is 
a vector field. If we parametrize positions on the $w$-complex plane as
$w=a+ib$ and $V = (V_a - iV_b)/2$, then $\vec{V} = \vec{\nabla}_w \Phi$.
The term "velocity" is related to the underlying Burgers dynamics to be
discussed later. To summarize, the limiting eigenvalue density and the
eigenvector correlation function can be calculated from the electrostatic
potential using eqs. (\ref{Poisson}) and (\ref{OV}), respectively, and 
taking the limit $w\rightarrow 0$ which project quaternions to the $z$-plane. 

It remains to show how to calculate the electrostatic potential $\Phi(z,w,\tau)$ (\ref{pot}). The standard method is based on enumeration of planar diagrams
as mentioned above. The main object in this method is the 
Green's function $\mathcal{G}$.

Here we propose an alternative approach which is based on
the diffusion equation in the quaternionic $(2+2)$-dimensional space in the direction perpendicular to the $z$-complex plane. The primary object in this calculation is an extended characteristic polynomial defined in the next section. 

\section{Averaged extended characteristic polynomial}
\label{aecp}

In order to calculate the potential (\ref{pot}) in the limit 
$N\rightarrow \infty$ we rewrite it as
\begin{align}
\label{pot2}
	\Phi(z,w,\tau) = 
	\frac{1}{N} \left < \log \det \left ( (z-X)(\bar{z} - X^\dagger) + |w|^2 \right ) \right >_\tau.
\end{align}
and define an associated object -- an effective potential
\begin{align}
\label{logdet}
\phi(z,w,\tau) = 
\frac{1}{N} \log \left < \det \left ( (z-X)(\bar{z}-X^\dagger) + |w|^2 \right ) \right >_\tau \equiv \frac{1}{N} \log D(z,w,\tau),
\end{align}
where 
\begin{align}
\label{ddef}
D(z,w,\tau) = \left < \det \left ( (z-X)(\bar{z}-X^\dagger) + |w|^2 \right ) \right >_\tau = \left < \det \left ( \begin{matrix} z - X & - \bar{w} \\
 w & \bar{z} - X^\dagger  
 \end{matrix} \right ) \right >_\tau .
\end{align}
We refer to $D(z,w,\tau)$ as to averaged extended characteristic polynomial (AECP).
For the Gaussian process (\ref{gp}) the determinant $D(z,w,\tau)$ self-averages
for $N\rightarrow \infty$ and $\Phi$ can be
replaced by $\phi$ in this limit. The advantage of using the latter is that
the averaged extended characteristic polynomial $D(z,w,\tau)$ \eqref{ddef}, 
which appears in the definition of $\phi = \frac{1}{N} \log D$  obeys a simple
diffusion equation with respect to the variable $w$
\begin{align}
\label{ddiff}
	\partial_\tau D(z,w,\tau) = \frac{1}{N} \partial_{w\bar{w}} D(z,w,\tau),
\end{align}
as shown in Appendix A. 
Note that, from the point of view of this equation,
$z$ is a dummy parameter. The  $z$-dependence appears only in  the initial
condition
\begin{align}
\label{init}
D(z,w,0) = D_{0}(z,w) = \det \left ( (z-X_0)(\bar{z} - X_0^\dagger) + |w|^2 \right ),
\end{align}
which is completely determined by the initial matrix $X_0$. In other words,
for each $z$ we have an independent diffusion in 
the perpendicular $w$-complex plane. The problem is therefore exactly integrable, and 
the  solution of the diffusion equation (\ref{ddiff}) reads
\be
\label{DD0}
D(z,w,\tau) = \frac{N}{\pi \tau} \int_C 
\exp\left(-N \frac{|w-w'|^2}{\tau}\right) D_0(z,w') \dd^2 w',
\ee
where $D_0(z,w')=D(z,w',0)$. The solution can be equivalently written as
\be
\phi(z,w,\tau) = \frac{1}{N} \log 
\frac{N}{\pi \tau} \int_C 
\exp N \left(\phi_0(z,w')-\frac{|w-w'|^2}{\tau}\right) \dd^2 w',
\ee
where $\phi_0(z,w')=\frac{1}{N}\log D_0(z,w')$. In the limit $N\rightarrow \infty$
the last equation assumes the form of the Hopf-Lax formula \cite{HOPFLAX}
\be
\label{hl}
\phi(z,w,\tau) = \max_{w'} 
\left(\phi_0(z,w')-\frac{|w-w'|^2}{\tau}\right) .
\ee
This equation describes the evolution of the electrostatic 
potential in the limit $N\rightarrow \infty$ for the given initial
configuration $\phi_0(z,w)$, so this equation solves our original problem. 

A few remarks are in order. The characteristic polynomial 
$D(z,w,\tau)$ and the potential $\phi(z,w,\tau)$ (\ref{logdet}) depend on 
$w$ only through the norm $|w|^2$. The diffusion preserves the spherical 
symmetry of these quantities in the $w$-complex plane, so it is convenient 
to rewrite these equations in the radial part $r$ of $w=r e^{i\alpha}$ skipping
the dependence on the phase $\alpha$. In particular
(\ref{DD0}) takes the form
\be
\label{ddiffsol}
D(z,r,\tau)= \frac{2N}{\tau}\int_{0}^{\infty} r' \exp{\left(-N\frac{r^2+r'^2}{\tau}\right)} I_{0}\left(\frac{2Nrr'}{\tau}\right)D_{0}(z,r')\dd r' ,
\ee 
and (\ref{hl}) simplifies to
\be
\label{hlr}
\phi(z,r,\tau) = \max_{r'} 
\left(\phi_0(z,r')-\frac{(r-r')^2}{\tau}\right) . 
\ee
The second remark is on the role of the parameter $w$. Originally it was introduced as a regulator to the expression for the potential
\eqref{pot} and eventually sent to zero. Here we promote $w$ to a full
complex variable and analyse the dynamics of the model on the entire $w$-complex plane. 
This approach allows one to trace not only eigenvalues but also eigenvectors
of the random matrix $X$ and to break the symmetry between matrices having 
identical eigenvalues but different eigenvectors. 

A complex valued matrix can be Schur decomposed $X = U (\Lambda + T) U^\dagger$ where $U$ is a unitary matrix, $\Lambda$ is a diagonal matrix containing the complex eigenvalues, and $T$ is a strictly upper-triangular matrix encoding information about eigenvectors. Two different matrices $X_1, X_2$ having the same eigenvalues $\Lambda$ but 
different eigenvectors have different $T_1$ and $T_2$.
The averaged extended characteristic polynomial \eqref{ddef} 
for these matrices differs $D_1(z,w) \neq D_2(z,w)$ when $|w| \neq 0$.
From this difference one can read off information about eigenvectors.

Moreover, the dynamics of the model in the quaternionic space has a beautiful 
physical interpretation in terms of the Burgers dynamics. The behavior of the
model on the $z$-complex plane is a shadow of this dynamics. In particular, the support of the eigenvalue density $\rho(z,\tau)$ coincides with the location of shocks of the quaternionic Burgers dynamics in the full quaternionic $(z,w)$-space.

\section{Burgers dynamics}
\label{pairburg}
In the previous section we have found a solution to the diffusion equation (\ref{ddiff}). Here for completeness we relate the diffusion equation to Burgers dynamics. Using the definition (\ref{logdet}) it is easy to see that the effective potential $\phi=\phi(z,w,\tau)$ and its gradient $v(z,w,\tau) = \partial_w \phi(z,w,\tau)$  fulfill the following differential equations
\begin{align}
\label{kpz}
	\partial_\tau \phi = \frac{1}{N} \partial_{w\bar{w}} \phi + 
	\partial_w \phi \partial_{\bar{w}} \phi,
\end{align}
and
\begin{align}
	\partial_\tau v = \frac{1}{N} \partial_{w\bar{w}} v + \partial_w |v|^2, \label{vequ}
\end{align}
respectively. These equations describe Burgers dynamics on the $w$-complex plane for a two dimensional velocity field $v=(V_a - iV_b)/2$ derived from the potential $\phi$: 
$\vec{V}= \vec{\nabla}_w \phi = (\partial_a \phi, \partial_b \phi)$ where $w=a+ib$.
The coefficients of the Laplacian term can be identified as a hydrodynamic viscosity parameter $\nu = \frac{1}{N}$. Equation (\ref{vequ}) follows from \eqref{ddiff} by an inverse Cole-Hopf transformation~\cite{COLEHOPF}. One can also write an equation for the $z$-gradient, $g(z,w,\tau) = \partial_z \phi(z,w,\tau)$:
\be
\label{gequ}
\partial_\tau g = \frac{1}{N} \partial_{w\bar{w}} g + \partial_z |v|^2. 
\ee
The two gradients are related to each other as $\partial_w g = \partial_z v$. The effective potential $\phi$ and the gradients reproduce the electrostatic potential and the quaternionic Green's function in the limit $N\rightarrow \infty$ 
\begin{align}
\label{phiPhi}
 \phi  \longrightarrow \Phi , \qquad 
 \left(\begin{array}{rr} g & v \\ -\bar{v} & \bar{g} \end{array} \right) \longrightarrow \mathcal{G}. 
\end{align}
Let us now discuss the inviscid limit $N\rightarrow \infty$. The effective potential (\ref{kpz}) obeys the equation 
\begin{align}
\label{kpzinfty}
	\partial_\tau \phi =  \partial_w \phi \partial_{\bar{w}} \phi.
\end{align}
which after applying rotational symmetry (in the variable $w$) simplifies to 
an equation
\begin{align}
\label{leq}
	\partial_\tau \phi = \frac{1}{4} (\partial_r \phi)^2 ,
\end{align}
for the radial variable $r=|w|$. The solution is given by
the Hopf-Lax formula (\ref{hlr}) which in our case is equivalent to
\begin{align}
\label{hlr2}
\phi(z,r,\tau) = \phi_{0}(z,r_*) - \frac{(r-r_*)^2}{\tau},
\end{align}
with $r_*$ being the location $r_*=r'$ of the maximum (\ref{hlr}) 
given by the usual extremum condition
\begin{align}
\label{rstar}
	\partial_r \phi_{0}(z,r_*) = \frac{2(r_*-r)}{\tau} .
\end{align}
To complete the scheme, the maximizing parameter $r_*$ has to be calculated from 
(\ref{rstar}) and the result $r_*=r_*(z,r,\tau)$ has to be inserted to (\ref{hlr2}).

We solve this equation for $\phi_{0}(z,r) = \frac{1}{N} \Tr \log \mathcal{M}(z,r)$ where $\mathcal{M}(z,r) = (z-X_0)(\bar{z}-X_0^\dagger) + r^2$ with an initial matrix $X_0$
\begin{align}
\label{rr}
\phi(z,r,\tau) = \frac{1}{N} \Tr \log \mathcal{M}(z,r_*) - \frac{(r-r_*)^2}{\tau}, \qquad \frac{r_*}{N} \Tr \mathcal{M}(z,r_*)^{-1} = \frac{r_*-r}{\tau},
\end{align}
Eliminating $r_*$ from this set of equation we obtain the effective potential $\phi(z,r,\tau)$ for an arbitrary initial matrix $X_0$. From it we derive 
the limiting density $\rho(z,\tau)$ \eqref{Poisson} and the eigenvector
correlations $O(z,\tau)$ \eqref{OV}
\begin{align} \label{po}
	\rho(z,\tau) & = \frac{1}{\pi} \partial_{z\bar{z}} \phi(z,0,\tau), \\
	O(z,\tau) & = \frac{1}{4\pi} \lim_{r\to 0} \left ( \partial_r \phi(z,r,\tau) \right )^2.
\end{align} 
For the initial condition of the form $\phi_{0}(z,r) = \frac{1}{N} \Tr \log \mathcal{M}(z,r)$ we arrive, after differentiation and some algebraic manipulations, at 
\begin{align}
\label{rhokpz}
	\rho(z,\tau) & = \frac{1}{N\pi} \frac{1}{\Tr\mathcal{M}^{-2} } \det \left ( \begin{matrix}  
	\Tr (\bar{z} - X_0^\dagger)\mathcal{M}^{-2} & \Tr \mathcal{M}^{-2} r_* \\
	-\Tr \mathcal{M}^{-2} r_* &\Tr (z - X_0)\mathcal{M}^{-2} \end{matrix} \right ) + \frac{1}{N\pi} \Tr \left (\mathcal{M}^{-1} [ \mathcal{M}^{-1}; z-X_0] (\bar{z} - X_0^\dagger) \right ), \\
	\label{okpz}
	O(z,\tau) & = \frac{1}{\pi \tau^2} r_*^2,
\end{align}
where $\mathcal{M} = \mathcal{M}(z,r_*)$. In the final formulas we set $r=0$
to project the results to the $z$-complex plane. The equation for $r_*$ (\ref{rr}) simplifies for $r=0$ to
\begin{align}
\label{bkpz}
	\frac{1}{N} \Tr \mathcal{M}(z,r_*)^{-1} = \frac{1}{\tau}.
\end{align}
Equations \eqref{rhokpz}, \eqref{okpz} are valid inside the boundary given by
\begin{align}
	\frac{1}{N} \Tr \mathcal{M}(z,0)^{-1} = \frac{1}{\tau},
\label{bckpz}
\end{align}
which corresponds to $r_*=0$. Outside this boundary $O(z,\tau)= 0$ and $\rho(z,\tau)=0$. For a normal initial matrix $X_0$, the second term in the spectral density \eqref{rhokpz} drops out since $[\mathcal{M}^{-1}, z-X_0] = 0$. 

By inspecting the boundary equation \eqref{bckpz} one finds a surprising connection to the so-called pseudospectrum \cite{TREFETHEN}, a mathematical concept generalizing the notion of the eigenvalue spectrum. Pseudospectrum of a matrix $A$ is defined as an subset $\sigma_A$ of the complex plane $z$ such that
\begin{align}
	|| (z-A)^{-1}|| > \frac{1}{\epsilon} ,
\end{align}
where the symbol $||\cdot ||$ is some arbitrary matrix norm and $\epsilon$ is the parameter of the pseudospectrum. In the $\epsilon \to 0$ limit, one recovers the standard eigenvalues as poles of the resolvent $(z-A)^{-1}$.  \\
For the initial matrix $A=X_0$, the boundary of the pseudospectrum subset $\sigma_{X_0}$ is exactly the eigenvalue boundary \eqref{bckpz} with $\epsilon^2 = \frac{\tau}{N}$ and a Frobenius norm $|| X ||_F = \sqrt{\Tr X^\dagger X}$. From this simple identification we conclude that a diffusion model with an initial matrix $X_0$ is also a probabilistic realization of the pseudospectrum for the same matrix.

We finish this section by discussing an equation for the gradient $v=\partial_w \phi$
in the inviscid limit. The equation is equivalent to the one for the potential $\phi$ that we discussed above but in some situations the equation for the gradient is 
more handy to use. The inviscid version of the
(\ref{vequ}) reads
\begin{align}
\label{burg2d}
	\partial_\tau v = \partial_w |v|^2.
\end{align}
It is an inviscid Burgers equation in $2+1$ dimensions. 
A general solution to this equation for smooth (differentiable) initial conditions can be deduced from the Hopf-Lax formula (\ref{hl}) for the effective potential $\phi$  for $N\rightarrow \infty$. The maximum in (\ref{hl}) is achieved for $w'=w_*$ fulfilling the standard extremum condition 
\be
0= \bar{v}_0(z,w_*) - \frac{w_*-w}{\tau},
\ee 
which is equivalent to 
\be
\label{wxi}
w = w_* -  \tau \bar{v}_0(z,w_*) ,
\ee 
for which  $v(z,w,\tau) = v_0(z,w_*)$.  Inserting $w_*$ (\ref{wxi}) to this equation
we get a solution to (\ref{burg2d}) 
\be
\label{2dsol}
v = v_0(z,w+\tau \bar{v}),
\ee
which is given by an implicit equation for $v=v(z,w,\tau)$ depending only on the initial condition $v_0(z,w)$. The parameter $w_*$ can be viewed as a labeling parameter for the family of the characteristic lines. These lines start to cross when the labeling fails to be bijective i.e. $\frac{dw}{dw_*} = 0$. This singularity condition defines a caustic surface - a boundary across which an ambiguity of the solutions arises. The development of multivalued solutions is an unwanted feature of the inviscid Burgers equation and is circumvented by constructing shock lines along which one cuts the characteristics rendering the solution unique.

The two-dimensional Burger's evolution (\ref{burg2d}) can be simplified in our case due to the rotational symmetry. We are interested only in the solutions which depend on the modulus $r=|w|$.  In this case the vector velocity field
is a central field $v = \frac{\bar{w}}{r} \nu$, with $\nu=|v|$, and the vector equation (\ref{burg2d}) reduces to a scalar equation for the modulus of the velocity field
\begin{align}
\label{vsym}
	\partial_\tau \nu = \nu \partial_r \nu,
\end{align} 
with a solution given by 
\begin{align}
\label{sol1d}
\nu = \nu_0(z,r+\tau \nu).	
\end{align}
From this solution we can reconstruct the full $2d$-solution:
$v(z,w,\tau) = \frac{\bar{w}}{r}\nu(z,r,\tau)$ with $r=|w|$.


\section{Examples}
\label{examples}
In this section we discuss three examples: (1) the canonical Ginibre evolution for which the initial matrix is $X_0 = 0$, (2) the spiric evolution for which $X_0 = \diag(-a,-a...,a,a...)$ with an equal number of $\pm a$ and (3) an
evolution initiated from a non-normal matrix: $(X_0)_{ij} = \alpha \delta_{i,j-1}$.
This matrix has eigenvalues equal zero as the initial matrix in the first example but it is not a normal matrix.

The first example serves as a proof-of-concept. We solve the Burgers equations by the method of characteristics to obtain the spectral density, the eigenvector correlator and the potential function in the large $N$ limit. The discussion is accompanied by Appendices B,C where we consider finite $w$ results and a relation of the averaged extended characteristic polynomial to the two-point kernel of the underlying determinantal process.

The second example illustrates an evolution of the eigenvalue density initiated from two disconnected eigenvalue "islands" which grow in the course of time to collide at some critical time. We discuss a novel universality arising in the vicinity of the collision. 

The third example demonstrates the dependence of the evolution of the eigenvalue distribution on the initial information which goes beyond the eigenvalues themselves.

\subsection{Ginibre evolution}
\begin{figure}[ht!]
	\centering
	\includegraphics[width=1.\textwidth]{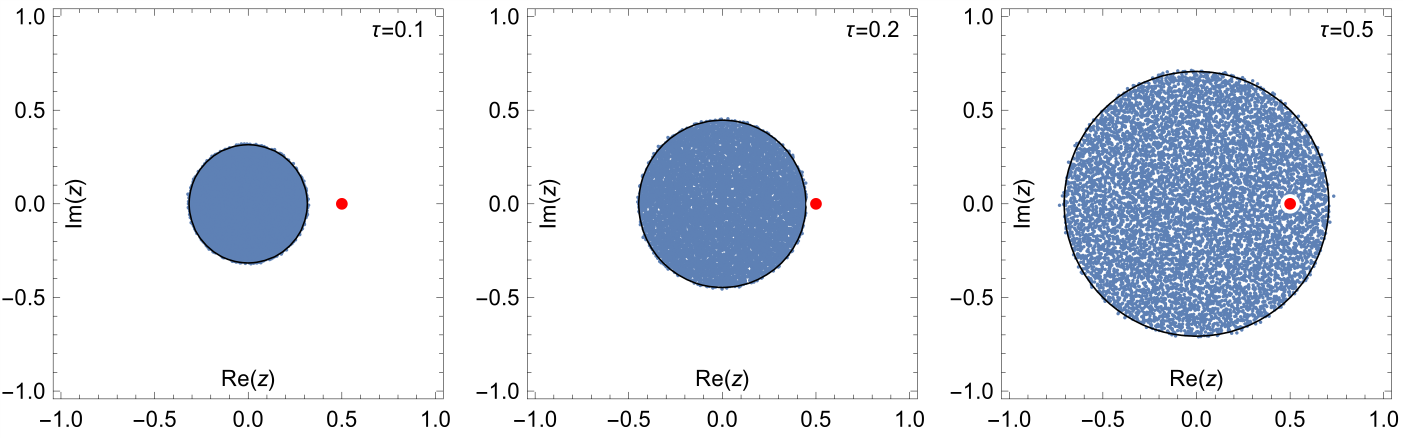}
      \caption{A numerical simulation of the spectral density with an initial matrix $X_0=0$ at time slices $\tau=0.1,\tau=0.2$ and $\tau=0.5$ respectively, an ensemble of $6$ matrices of size $N=1500$. Black curves are the large $N$ spectral boundaries and the red dot indicates an arbitrary spectator position $z$ where the evolution is probed.}
      \label{numgg}
\end{figure}

\begin{figure}[ht!]
  \centering
   \includegraphics[scale=.55]{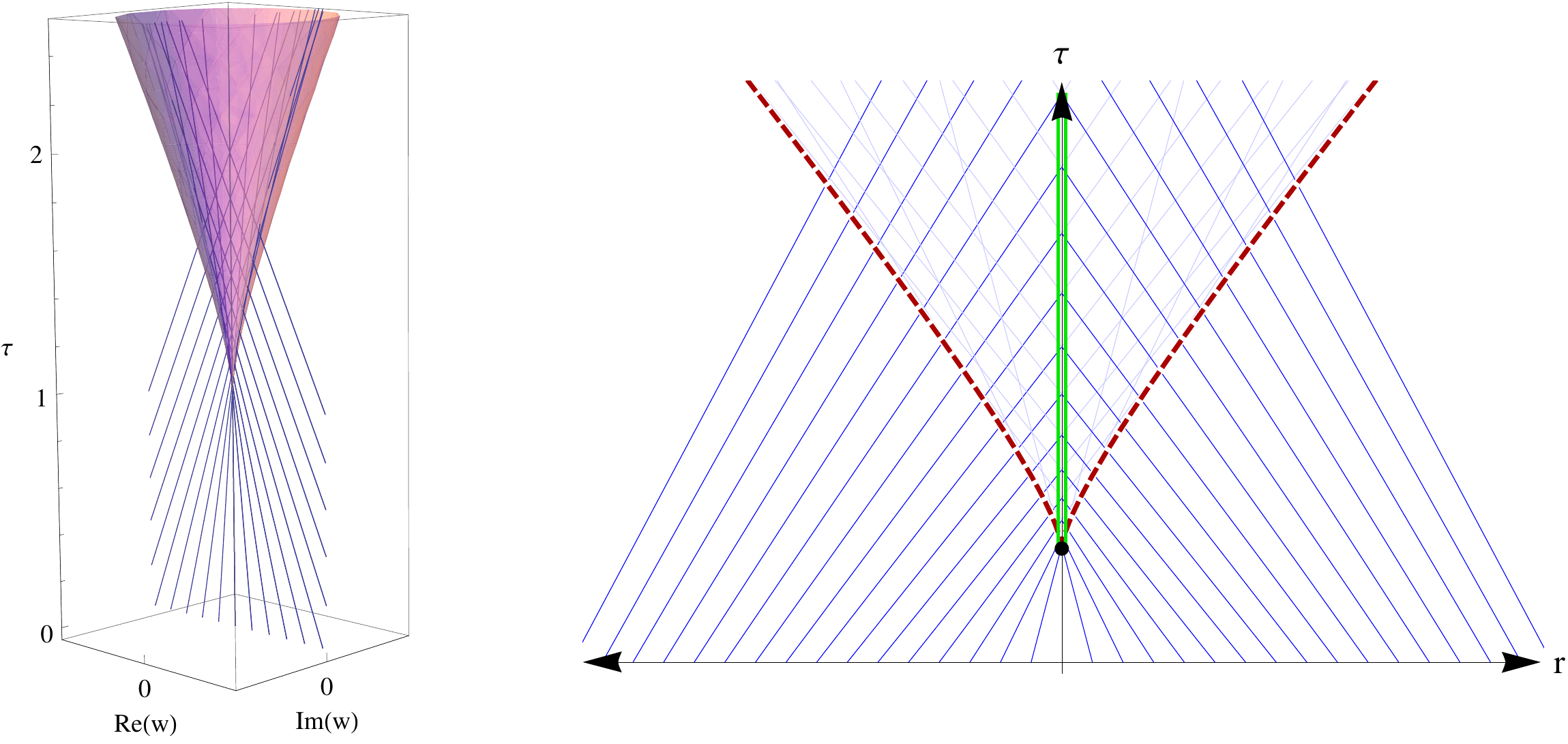}
      \caption{The characteristic lines at $z=1$ for the vector 2+1 Burgers equation
      (left) and a 1+1 section (right). The caustic cone-like surface on the left plot is denoted by dashed line on the right one. The shock is located on the cones' axis
It start from its apex. It is shown a green double line on the vertical axis in the right chart. The vertical line corresponds to $r=0$, that is to the place where the quaternionic pair $(z,w)$ reduces to $(z,0)$ that lies on the $z$-complex plane.}
      \label{charsgg}
\end{figure}

The evolution is initiated from the matrix $X_0 = 0$. The evolution of the eigenvalue density is shown in Fig. \ref{numgg}. The spectral density forms a circular eigenvalue "island" expanding in time. For each time $\tau$ the ensemble of matrices
in this evolution is equivalent to a Ginibre Ensemble with a time-rescaled dispersion.
The initial condition for the determinant (\ref{init}) is $D_0(z,w)=\left(|z|^2+|w|^2\right)^N$, for the effective potential (\ref{logdet}) $\phi_0(z,w)=\log (|z|^2+|w|^2)$, for the velocity $v_0(z,w)=\bar{w}/(|z|^2+|w|^2)$, and for its modulus $\nu_0(z,r) = r/(|z|^2+r^2)$, where $r=|w|$, respectively. We solve
the inviscid 2+1 Burgers equation \eqref{burg2d} using the method of characteristics.
Characteristics \eqref{wxi} for this equation are shown in 
Fig. \ref{charsgg}. In the left panel we show a plot in 2+1 dimensions and
in the right one its 1+1 dimensional section. 
The evolution takes place in the $w$ complex plane however the position on $z$-plane
defines the initial condition. The $z$-variable acts as a spectator on the eigenvalue plane "observing" the evolution in the perpendicular $w$-direction. We identify the cone-like caustic surface (left plot) whose apex is located at $r=0$ and at critical time $\tau_c = |z|^2$. This surface is the boundary along which the characteristics start to cross, making the Burgers solution multi-valued. This ambiguity develops for $\tau>\tau_c$ as the expanding eigenvalue boundary "swallows" the spectator at $z_0$. We find the position of the shock line as a locus distanced equally from the caustic surface at each given time (the Rankine-Hugoniot condition). Since our problem is radially symmetric, the shock is positioned exactly at $r=0$ and starts from the critical time $\tau_c$. Therefore although the dynamics takes place in the whole $r$ space, the shock is always confined to the "physical" $r=0$ region. Moreover if we confine it to $r=0$, the spectator at $z$ stays on the shock line at every time $\tau> \tau_c$. We conclude that inside the bulk of the spectrum (i.e. the non-holomorphic sector), the observer is constantly on the shock line.

\begin{figure}[ht!]
  \centering
   \includegraphics[scale=.9]{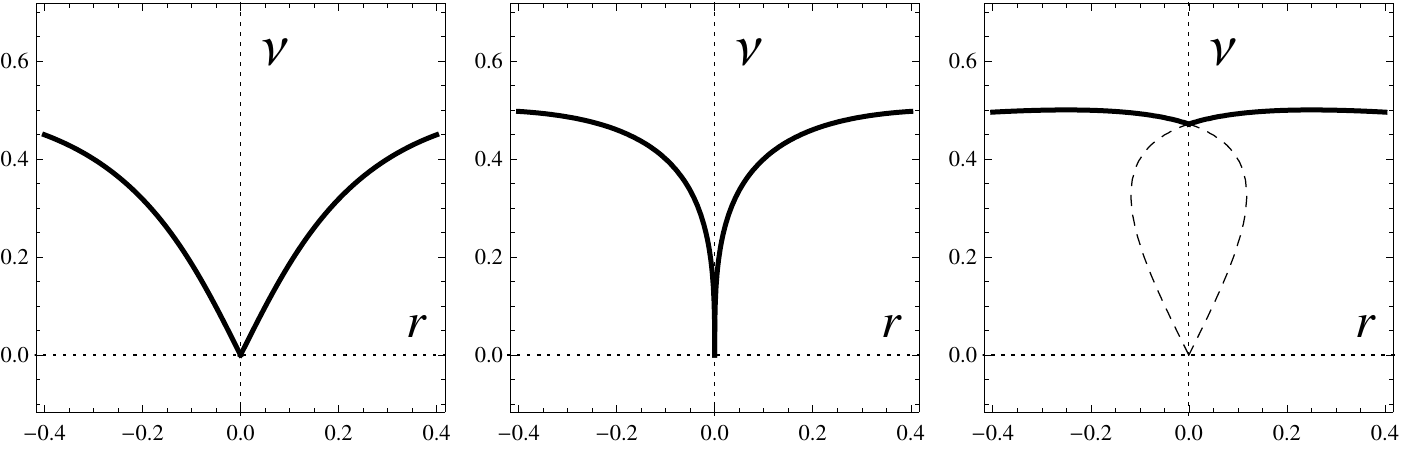}
      \caption{The solution $|v_\tau|$ for fixed $z_0=1$ and three different times $\tau=0.5,\tau=1$ and $\tau=1.5$ respectively. It shows the development of nonzero solution for $\tau>\tau_c$ and the emergence of an additional unphysical solution depicted by dashed lines.}
      \label{vsolgg}
\end{figure}

Having identified the positions of the shock we can write down a solution of 
\eqref{sol1d} for the reduced 1+1 Burgers equation for our initial conditions 
$\nu_0(z,r) = r/(|z|^2+r^2)$. It reads
\begin{align}
	\nu = \frac{ r + \tau \nu}{(r + \tau \nu)^2 + |z|^2} .
\end{align}
It is an implicit algebraic equation for $\nu=\nu(z,w,\tau)$. It can be rewritten as
a cubic equation. Solutions for different $\tau$ is plotted in Fig. \ref{vsolgg}. Rather
than showing the solution for the modulus $\nu$ we show a cross section of the vector field which has two symmetric branches $\pm \nu$. The 
right plot shows the solution inside the caustic surface with an unphysical branch depicted as dashed lines. For $r=0$, that is on the  $z$-plane, we have 
\begin{align}
\label{vtauggsol}
	\nu(z,\tau)=\left\{ \begin{array}{ccc} 
	0 & {\rm for} & \tau < |z|^2 \\ 
	\frac{1}{\tau} \sqrt{\tau-|z|^2} & {\rm for} & \tau > |z|^2 
	\end{array}\right. .
\end{align}
The boundary $|z|^2 = \tau$ is found by the sewing condition of zero and non-zero solutions. 
From $\nu$ we readily obtain the eigenvector correlation function \eqref{evs} in the large $N$ limit 
\begin{align}
\label{o}
O(z,\tau) = \frac{1}{\pi} \nu^2 = \left\{ \begin{array}{ccc} 0 & {\rm for} &  \tau < |z|^2  \\ \frac{1}{\tau^2 \pi} (\tau - |z|^2)  & {\rm for} & \tau > |z|^2   \end{array}\right. .
\end{align}
We can also find the second gradient $g=\partial_z \phi$ of the effective potential.
To this end we can use the equation (\ref{gequ}) which in the limit $N\rightarrow \infty$ simplifies to $\partial_\tau g = \partial_w |v|^2$. For our initial condition $g_{0}(z,r) = \bar{z}/(|z|^2+r^2)$ it gives
\be
g(z,\tau)=\left\{ \begin{array}{ccc} 1/z & {\rm for} & \tau<|z|^2 \\
 \bar{z}/\tau & {\rm for} & \tau>|z|^2 \end{array}\right. .
\ee
Now we can use the Gauss law \eqref{gausslaw} to obtain the limiting eigenvalue 
density
\begin{align}
\rho(z,\tau) = \frac{1}{\pi\tau} \theta(\sqrt{\tau}-|z|) .
\label{Gi1}
\end{align}
It is given by a uniform distribution on a disc of radius $\sqrt{\tau}$.  For $\tau=1$, results (\ref{o},\ref{Gi1})
reproduce results of~\cite{CHALKERMEHLIG,GG}, respectively. We could alternatively find the same formulas by directly applying \eqref{rhokpz} and \eqref{okpz}. The maximizer $r_*$ can be found by solving the constraint \eqref{bkpz}:
\begin{align}
	r_* = \left\{ \begin{array}{ccc} 0 & {\rm for} & \tau<|z|^2 \\
 \sqrt{\tau-|z|^2} & {\rm for} & \tau>|z|^2 \end{array}\right. ,
\end{align}
We find
\begin{align}
\phi(z,\tau) = \left\{ \begin{array}{ccc} \ln |z|^2 & {\rm for} & \tau<|z|^2 \\
  \ln \tau +\frac{|z|^2}{\tau} - 1 & {\rm for} & \tau>|z|^2 \end{array}\right. .
\end{align}
The last formula was obtained for $\tau = 1$ in~\cite{KHORFYOD}. 

So far we have discussed the limit $N\rightarrow \infty$, in which as the effective potential $\phi$ and its gradients are equal to the electrostatic potential
and the quaternionic Green's function (\ref{phiPhi}). What about finite $N$?
For finite $N$ the order of calculating the average and the logarithm in (\ref{pot2})
and (\ref{logdet}) matters , so $\phi$ is at best only an approximation of $\Phi$
for large but finite $N$. So clearly our method is not able to get insight into the finite
$N$ corrections. Surprisingly as we show below, the behavior of the diffusion 
kernel $D(z,w,\tau)$ on the $z$ complex plane, that is for $w=0$ reveals the same type
of finite size effects as known from exact calculations of the eigenvalue density for
Ginibre ensemble \cite{CHALKERMEHLIG}. For our initial conditions the diffusion kernel (\ref{ddiffsol}) is 
\begin{align}
D(z,r,\tau)= \frac{2N}{\tau}\int_{0}^{\infty} r' \exp{\left(-N\frac{r^2+r'^2}{\tau}\right)} I_{0}\left(\frac{2Nrr'}{\tau}\right)(|z|^2+r'^2)^N \dd r' 
\end{align}
On the $z$-complex plane that is for $r=0$ this integral significantly simplifies and it 
can be for large $N$ calculated using the  saddle point method. There are three saddle points
\begin{align}
	r'_0 = 0, r'_\pm = \pm \sqrt{\tau-|z|^2}.
\end{align}
When $\tau$ approaches $|z|^2$ from above, the two points $r'_\pm$ approach each other along the real axis to coalesce for $\tau=|z|^2$ at $r'=0$. When $\tau$ becomes smaller
than $|z|^2$ the two points move on the imaginary axis. Near the critical value
$\tau=|z|^2$ it is convenient to introduce a rescaled parameters
\begin{align}
	r' & = \theta N^{-1/4}, \qquad |z| = \sqrt{\tau} + \eta N^{-1/2}
\end{align}
and use them in the calculations. One finds that for large $N$ the diffusion kernel
behaves as
\begin{align}
D(z=\sqrt{\tau} + \eta N^{-1/2},r=0,\tau) 
\sim \frac{1}{2\pi\tau} \text{erfc} \left ( \sqrt{\frac{2}{\tau}} \eta \right),
\end{align}
where  $\text{erfc}$ is the complementary error function. This result holds not
only on the $z$-complex plane but also sufficiently close to the $z$-complex plane
that is for $r$ approaching zero as $N^{-3/4}$ or faster: $r = O(N^{-3/4})$. Indeed in 
this case the argument of the function $I_0$ is a finite number and this function behaves
as a constant for large $N$.  
The error function behavior has the same form as the finite size expression for the eigenvalue density function. This indicates that there may be some deeper relation between the diffusion kernel (averaged extended characteristic polynomial) and the two-point kernel known from the considerations of the underlying determinantal process for Ginibre matrices. We work out this analogy at heuristic level in Appendices B,C.  

\subsection{Spiric case}
\begin{figure}[ht!]
  \centering
	\includegraphics[width=1.\textwidth]{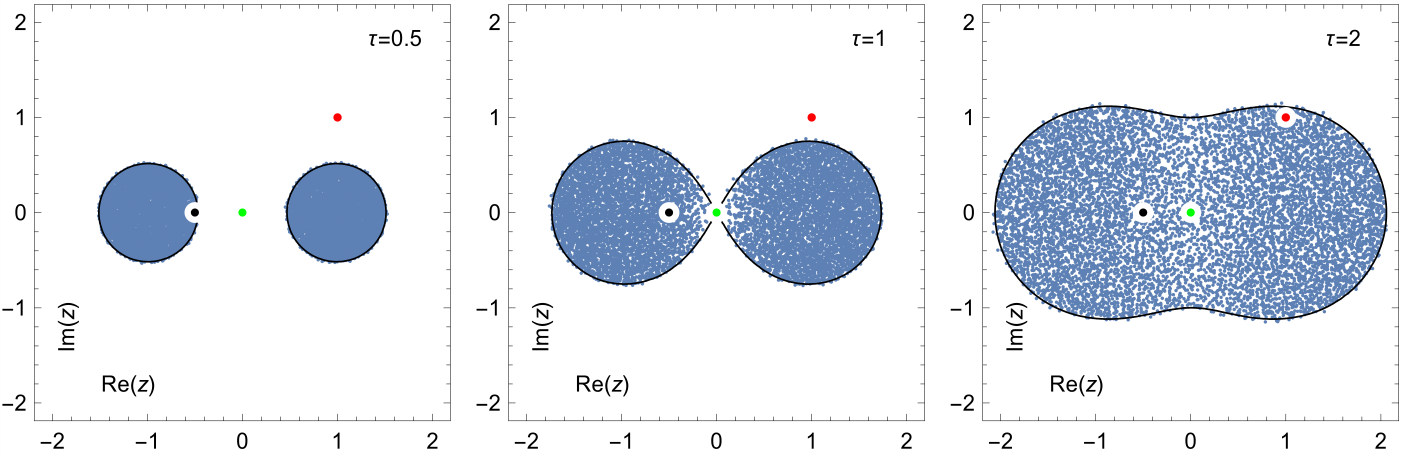}
      \caption{The spectral density dynamics for an initial matrix $X_0 = \text{diag}(1,...,1,-1,...,-1)$ before, at and after the critical time $\tau_c=1$, the ensemble consisted of $6$ matrices of size $N=1500$. The green, red and black dots denote three observers useful in the analysis of the evolution.}
      \label{spiricdyn}
\end{figure}

\begin{figure}[ht!]
  \centering
	\includegraphics[width=1.\textwidth]{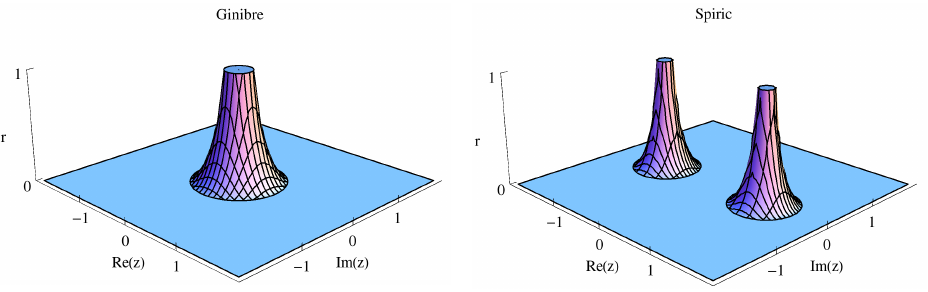}
      \caption{Caustic surface in the variables $(r,z)$ for Ginibre case (left plot) and spiric case (right plot) with time fixed at $\tau_0=0.5$.}
      \label{ggvsspirichats}
\end{figure}

We now consider diffusion initiated from a diagonal matrix $X_0 = \text{diag}(a,...,a,-a,...,-a)$ with the same 
number of $a$ and $-a$ eigenvalues. In the course of time two initial eigenvalue "islands", initially concentrated around $\pm a$, expand to collide at some critical time $\tau_c$ as shown in Fig. \ref{spiricdyn}.
The initial condition corresponds to $$D_0(z,r) = (r^2+|z-a|^2)^{\frac{N}{2}}(r^2+|z+a|^2)^{\frac{N}{2}},$$
or if we write it for the modulus of the velocity field \eqref{vsym}
\be
\nu_0(z,r)= \frac{1}{2} \frac{r}{r^2 + |z-a|^2} + \frac{1}{2} \frac{r}{r^2 +|z+a|^2}.
\ee
\begin{figure}[ht!]
  \centering
	\includegraphics[width=1.\textwidth]{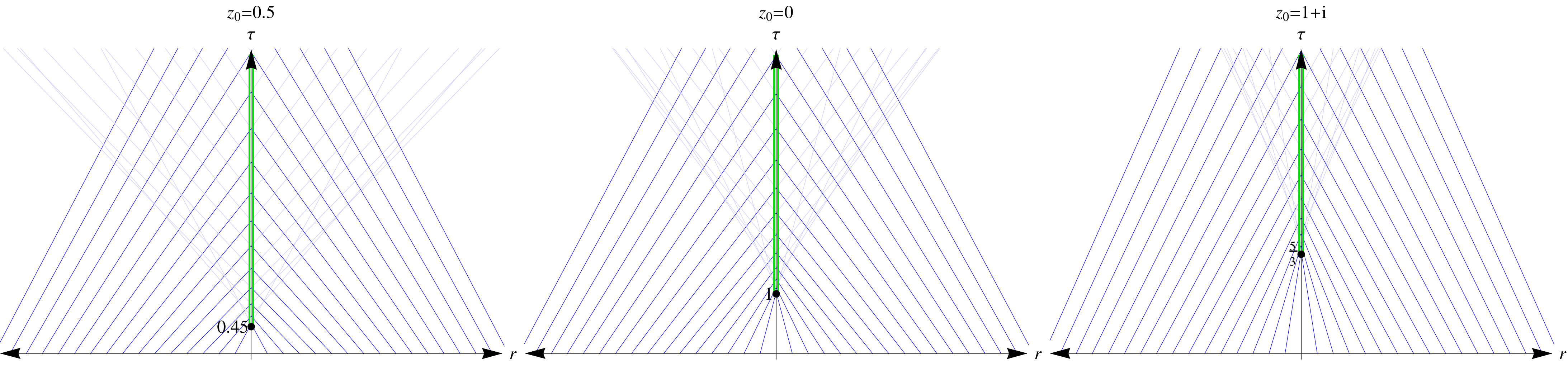}
      \caption{Characteristic lines in the spiric case for fixed observers $z_0=0.5$ (black dot on Fig. \ref{spiricdyn}), $z_0=0$ ((green dot on Fig. \ref{spiricdyn}) and $z_0=1+i$ (red dot on Fig. \ref{spiricdyn}) with $a=1$. From left to right, the shock formation occurs later in time which corresponds to meetings between observers $z_0$ and the expanding spectral boundary. \label{spiricchars}}
\end{figure}
In Fig. \ref{spiricchars} we show charecteristics for three different values of the observer position $z$, including $z=0$ where the collision takes place. As we can see the dynamics in the $r$ plane behaves qualitatively in the same way as for the Ginibre case. In particular , characteristics form the same type of caustic surfaces with a shock line present after some critical time $\tau_c = \frac{|a^2-z^2|^2}{|a|^2+|z|^2}$. This critical time can be found by sewing the solutions $\nu$ or by finding the position of the caustic surface apex as a function of $z$ and $\tau$. In Fig. \ref{ggvsspirichats} we show a comparison
of the caustic surfaces for Ginibre and spiric evolution for $\tau=0.5$.
For this inititial condition the solution \eqref{sol1d} of the inviscid Burgers equation 
takes the form
\begin{align}
	2 \nu =  \frac{r+\tau\nu}{|z+a|^2+(r+\tau\nu)^2} + \frac{r+\tau\nu}{|z-a|^2+(r+\tau\nu)^2} .
\end{align}
For $r=0$ the solution can be written explicitly
\begin{align}
\label{vtaussol}
	\nu(z,r=0;a)=\left\{ \begin{array}{ccc} 
	0 & {\rm for} & z \notin \mathcal{S} \\ 
	\frac{1}{\tau} \sqrt{\frac{\tau}{2} - |a|^2-|z|^2 + \frac{1}{2} S_a} & {\rm for} & z \in \mathcal{S}
	\end{array}\right. ,
\end{align}
where $S_a = \sqrt{\tau^2 + 4Z_a^2}$, $Z_a=\bar{z}a+z\bar{a}$. The symbol
$\mathcal{S}$ stands for the interior of the spiric section defined by the contour
\begin{align}
\label{quartic}
	\tau(|a|^2 + |z|^2) = |a^2 - z^2|^2.
\end{align}
Generally spiric section is a curve obtained by intersection of a torus and a plane parallel to its rotational symmetry axis. We plotted spiric curves as contours around the scatter plots for eigenvalue densities in Fig. \ref{spiricdyn}. The eigenvector correlation function inside the spiric section $\mathcal{S}$ is 
\begin{align}
\label{spiriceigenvec}
	O(z,\tau) = \frac{\nu^2}{\pi}= \frac{N}{\tau^2 \pi} \left ( \frac{\tau}{2} -|a|^2 -|z|^2 + \frac{1}{2} S_a \right ).
\end{align}
We can also calculate the diagonal element of the Green's function using the equation
(\ref{gequ}) that in the limit $N\rightarrow \infty$ simplifies to $\partial_\tau g = \partial_z \nu^2$. For the initial condition as $g_{0} = \frac{1}{2} \left ( \frac{\bar{z} + \bar{a}}{r^2 + |z+a|^2} + \frac{\bar{z} - \bar{a}}{r^2 + |z-a|^2} \right )$ the 
solution reads
\be
g(z,r=0,\tau)=\left\{ \begin{array}{ccc} 
\frac{z}{z^2-a^2} & \text{for} \quad z \notin \mathcal{S} \\ 
\frac{\bar{z}}{\tau} -\frac{\bar{a} S_a}{2\tau Z_a} + c(z,a) & \text{for} \quad z \in \mathcal{S} \end{array}\right. ,
\ee
with a constant $c(z,a) = \frac{\bar{a}}{2Z_a}$ obtained by the sewing condition along the spiric section $\mathcal{S}$. We use the Gauss law to obtain the spectral density as
\begin{align}
\label{densityspiric}
	\rho(z,\tau) = \frac{S_a (2 - \tau |a|^2) + \tau^2 |a|^2}{2\pi \tau Z_a^2 S_a}.
\end{align}
The same results can be obtained from the calculations of the effective potential $\phi$
using the equations \eqref{rhokpz}, \eqref{okpz} and \eqref{bkpz} for $r_*$:
\begin{align}
	r_*= \left\{ \begin{array}{ccc} 0 & {\rm for} & z \notin \mathcal{S} \\
 \sqrt{\frac{\tau}{2} - |a|^2-|z|^2 + \frac{1}{2} S_a}& {\rm for} & z \in \mathcal{S} \end{array}\right. .
\end{align}
Outside $\mathcal{S}$ we get
\begin{align}
	\phi^{out}(z,\tau) = \frac{1}{2} \ln |z-a|^2 + \frac{1}{2} \ln |z+a|^2 ,
\end{align}
and inside $\mathcal{S}$
\begin{align}
 \phi^{in}(z,\tau) = \frac{1}{2} \ln \left (\frac{\tau^2}{2} + \frac{\tau}{2}S_a \right ) + \frac{1}{\tau}\left ( |a|^2 + |z|^2 -\frac{\tau}{2} - \frac{1}{2} S_a \right ) .
\end{align}
We recall that $S_a = \sqrt{\tau^2 + 4Z_a^2}$, $Z_a=\bar{z}a+z\bar{a}$, as defined after (\ref{vtaussol}).
\begin{figure}[ht!]
  \centering
	\includegraphics[width=1.\textwidth]{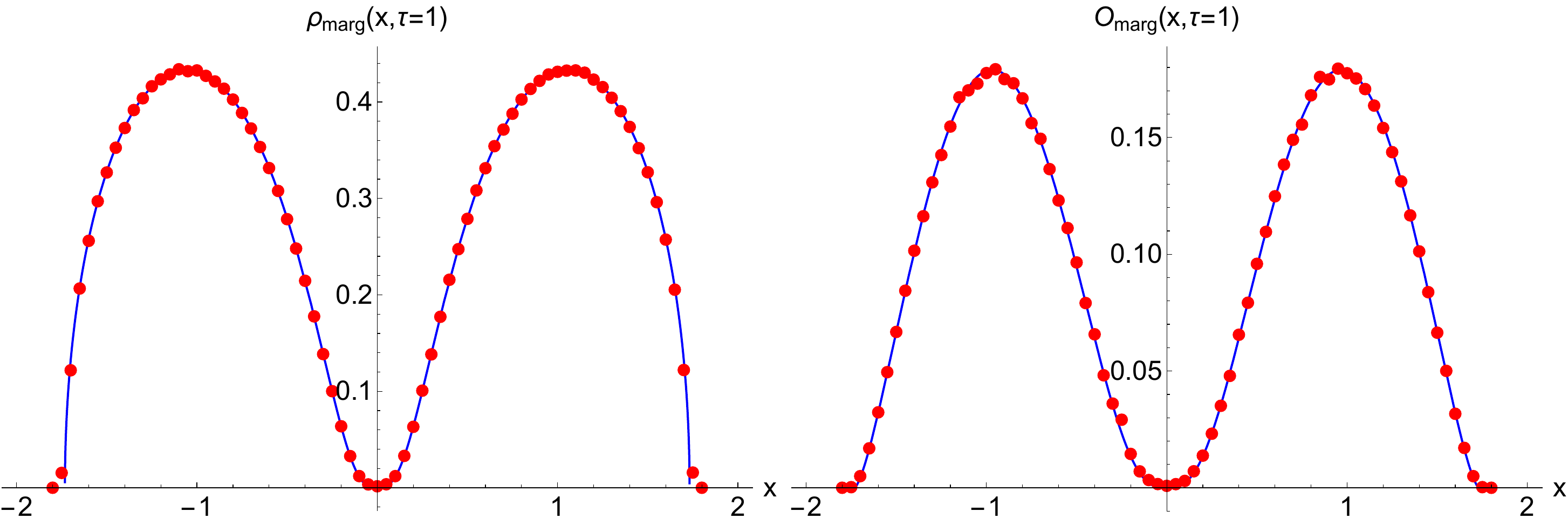}
      \caption{Numerical results in the spiric case for $a=1$ and critical time $\tau=1$ for eigenvector correlator (right plot) and spectral density (left plot) averaged over the imaginary axis, the ensemble consisted of $6 \cdot 10^3$ matrices of size $N=1000$.
      \label{spiricdennum}}
\end{figure}
In Fig. \ref{spiricdennum} we show a comparison of theoretical predictions for the limiting eigenvalue density \eqref{densityspiric} and the eigenvector correlation function \eqref{spiriceigenvec} with Monte-Carlo simulations. The agreement is very good.

\begin{figure}[ht!]
  \centering
	\includegraphics[width=1.\textwidth]{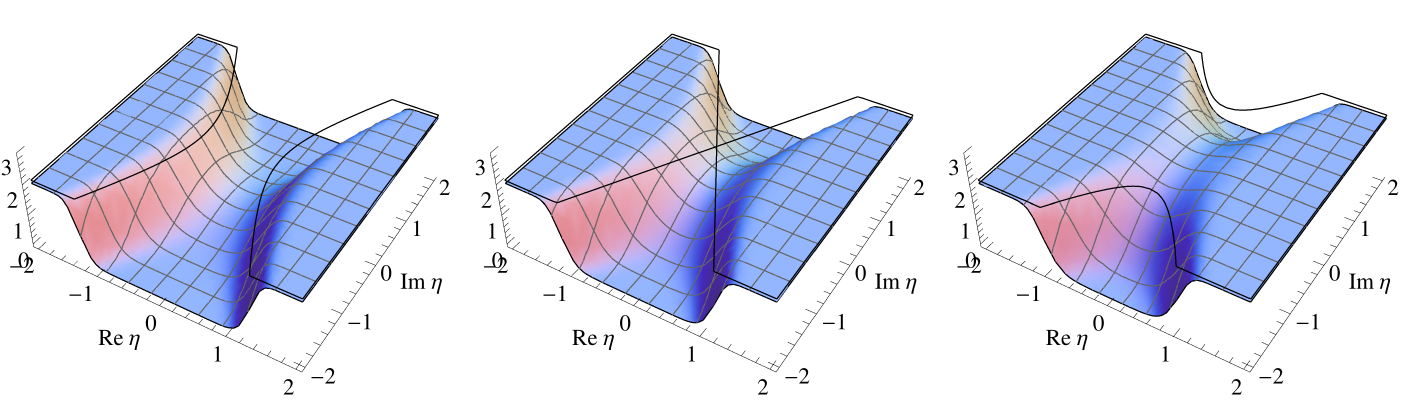}
      \caption{Evolution of the characteristic polynomial in the vicinity of the collision for rescaled times $t = -1, t=0,t=1$. Black contour on top is the large $N$ boundary of non-zero spectral density obtained from \eqref{quartic}.}
      \label{spiricuniv}
\end{figure}

We complete  the discussion of the spiric case by deriving a finite $N$ formula for the diffusion kernel $D(z,r,\tau)$ 
\be
\label{solspiric}
D(z,r,\tau)= \int_{0}^{\infty}r' \exp{\left(-N\frac{r^2+r'^2}{\tau}\right)} I_{0}\left(\frac{2Nrr'}{\tau}\right)(r'^2+|z-a|^2)^{\frac{N}{2}}(r'^2+|z+a|^2)^{\frac{N}{2}}  {\rm d}r'.
\ee
We are interested in the behavior close to the origin $z=0$, $r=0$ 
for $\tau$ close to the collision time. Without loss of generality we perform calculations
for $a=1$. In this case the collision time is $\tau=1$. We set $r=0$ and zoom into the vicinity of the critical region
\begin{align}
	r' = \theta N^{-1/4}, \qquad z = \eta N^{-1/4}, \qquad \tau = 1 + t N^{-1/2} 
\end{align}
where the saddle points merge. After expanding the solution \eqref{solspiric} we obtain an asymptotic formula
\begin{align}
D(z = \eta N^{-1/4}, r=0,  \tau = 1 + t N^{-1/2}) \sim \sqrt{\frac{\pi}{128 N}} e^{- \frac{\sqrt{N}}{2} ((\eta + \bar{\eta})^2 - 2 |\eta|^2) } (\eta + \bar{\eta})^4 \text{erfc} \left [\frac{1}{\sqrt{2}} \left ( |\eta|^2 - (\eta + \bar{\eta})^2 - t \right ) \right ]. 
\end{align}
which is plotted in Fig. \ref{spiricuniv}. Again, the result holds not only for $r=0$ but
more generally for $r=O(N^{-3/4})$.


\subsection{Non-normal Ginibre case}
We consider now diffusion initiated from a matrix $(X_0)_{ij} = \alpha \delta_{i,j-1}$. The matrix $X_0$ has all eigenvalues equal to $0$.
\begin{figure}[ht!]
  \centering
	\includegraphics[width=1.\textwidth]{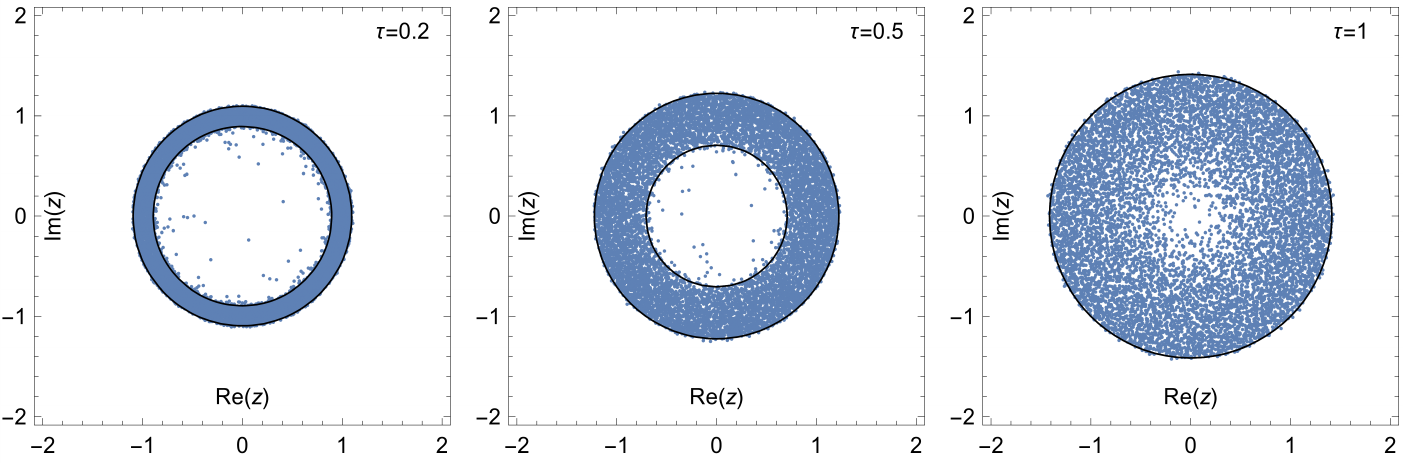}
      \caption{Evolution of the spectral density for non-normal initial condition $(X_0)_{ij} = \delta_{i,j-1}$, with time snapshots at $\tau=0.2$, $\tau=0.5$ and $\tau=1$, the ensemble consisted of $6$ matrices of size $N=1500$.}
      \label{nonnum}
\end{figure}
The initial eigenvalue distribution coincides with the one for the Ginibre case but afterwards the eigenvalue density obeys completely different evolution. Three snapshots of this evolution are shown in Fig. \ref{nonnum}. The initial distribution concentrated initially at zero 
instantaneously expands to a circle of radius $|\alpha|$, which corresponds to the pseudospectrum of the matrix. In the course of evolution  the support of the density takes the form of the  growing
annulus. After a finite time $\tau=|\alpha|^2$ the inner radius of the annulus shrinks to zero and eigenvalues fill up a full disk.

\begin{figure}[ht!]
  \centering
	\includegraphics[width=1.\textwidth]{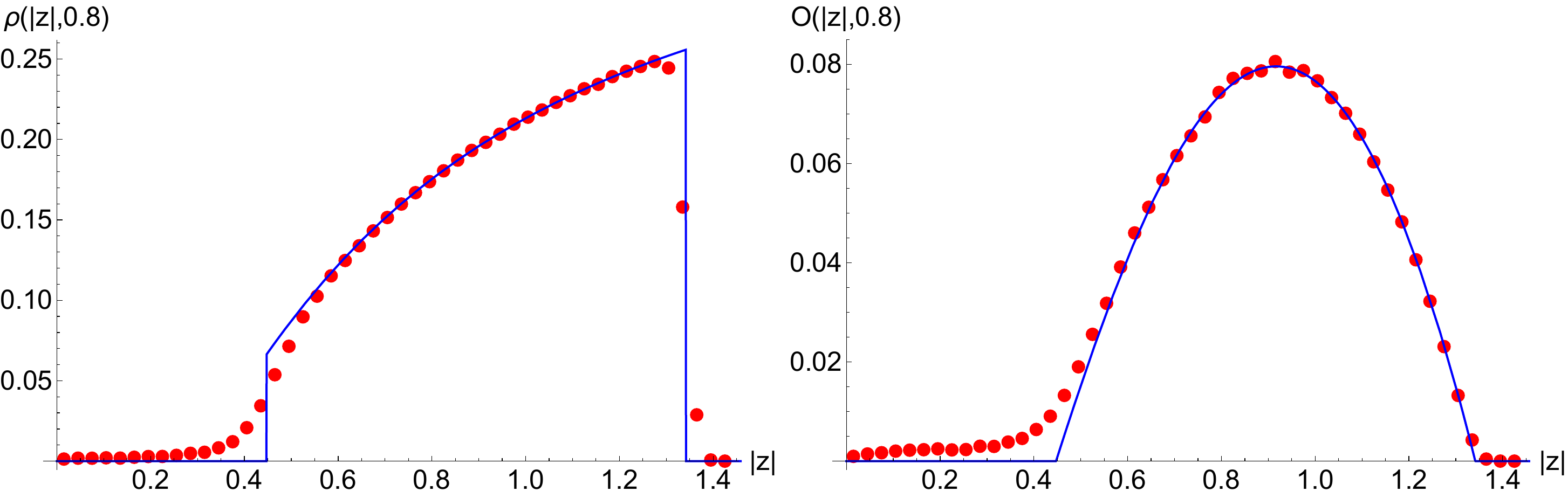}
      \caption{Numerical analysis of the eigenvector correlator (right plot) and the spectral density (left plot) for the non-normal initial condition, the ensemble consisted of $3\cdot 10^3$ matrices of size $N=1000$ with parameters $\alpha=1$ and time $\tau=0.8$.}
      \label{nonnum2}
\end{figure}

Let us show this by direct calculations. The matrix $\mathcal{M} = (z-X_0)(z-X_0)^\dagger + r^2$ (\ref{rr}) has for our choice of $X_0$ a tridiagonal form
\begin{align}
	\mathcal{M} = \left ( 
	\begin{matrix}
	a & b & 0 &  \hdots & 0 \\
	\bar{b} & a & b &   & 0\\
	\vdots & \bar{b} & \ddots & b & \vdots \\
	0 &  & \bar{b} & a & b \\
	0 & \hdots & 0 & \bar{b} & d \\
	\end{matrix} 
	\right ) 
\end{align}
with $a = |z|^2 + r^2 + |\alpha|^2$, $b = - \bar{z} \alpha$ and $d = |z|^2 + r^2$. The determinant of this matrix can be calculated explicitly: $\det \mathcal{M} = \frac{1}{\Delta}\left ( d(a_+^N - a_-^N) - |b|^2(a_+^{N-1} - a_-^{N-1}\right)$,
where $\Delta = \sqrt{a^2 - 4 |b|^2}$ and $a_{\pm} = \frac{1}{2} \left ( a \pm \Delta \right )$. The initial effective potential for large $N$ is
\be
\label{phia}
\phi_{0}(z,r) = \frac{1}{N} \Tr \log \mathcal{M} \approx \ln a_+ =
\ln \frac{1}{2}\left(a + \sqrt{a^2 - 4|z|^2|\alpha|^2}\right)  
\ee
where $a = |z|^2 + r^2 + |\alpha|^2$. We have neglected $1/N$ terms which
disappear in the limit $N\rightarrow \infty$. The value $r_*$ which maximizes the expression in the Hopf-Lax formula (\ref{hlr}) can be calculated from the equation (\ref{rstar}). For $r=0$ we get  
\begin{align}
	r_* = \left\{ \begin{array}{ccc} 0 & {\rm for} & z \notin \mathcal{A} \\
 \sqrt{T_\alpha-|z|^2 - |\alpha|^2 } & {\rm for} & z \in \mathcal{A} \end{array}\right. ,
\end{align}
where we used the notation $T_\alpha = \sqrt{\tau^2 + 4 |\alpha|^2 |z|^2}$. 
The boundary of the annulus $\mathcal{A}$ is given by the radii
\begin{align}
	|z|_\pm = \sqrt{|\alpha|^2 \pm \tau}.
\end{align}
We see that at the beginning of the evolution, $\tau\approx 0$, the annulus is infinitely narrow forming a one-dimensional circular pseudospectrum. On the other hand for
$\tau\rightarrow |\alpha|^2$ the inner radius tends to zero and the annulus becomes
a disk. Inserting $r_*$ to (\ref{hlr}) we obtain the potential 
\begin{align}
\phi(z,r=0,\tau) = \left\{ \begin{array}{ccl} \ln |\alpha|^2 & {\rm for} & |z| \in \left ( 0 ; |z|_- \right ) \\
  \ln \left ( \frac{\tau}{2} + \frac{1}{2}T_\alpha \right ) + \frac{|z|^2+|\alpha|^2}{\tau} - \frac{T_\alpha}{\tau} & {\rm for} & |z| \in \left ( |z|_- ; |z|_+ \right ) \\
  \ln |z|^2 & {\rm for} & |z| \in \left ( |z|_+ ; \infty \right ) 
   \end{array}\right. ,
\end{align}
for the three regions of the annulus. The spectral density is obtained from
the Poisson equation (\ref{po}) by differentiation the effective potential twice
\begin{align}
\label{nonherm}
	\rho(z,\tau) = \frac{1}{\pi \tau} \left ( 1 - \frac{|\alpha|^2}{T_\alpha} \right ) 
\end{align}
on the support of the annulus and zero otherwise.
Using \eqref{okpz} we obtain the eigenvector correlation function
 to the support of the annulus
\begin{align}
O(z,\tau) = \frac{1}{\pi \tau^2} \left (T_\alpha -|z|^2 - |\alpha|^2 \right ).
\end{align}
In Fig. \ref{nonnum2} we compared the prediction of the two formulas given above
with numerical simulations. The agreement is good. Deviations are observed
only close to the boundaries and can be attributed to finite-size effects.
It is easy to check that all the expressions transform to the corresponding expressions for the Ginibre evolution when $\alpha\rightarrow 0$.

Let us shortly discuss finite $N$ effects for the averaged extended characteristic polynomial. We are looking for an universal behaviour near the origin for $\tau \to 0$.
Around this space-time point an instantaneous transition from $N$ eigenvalues positioned at the origin to the ring of radius $|\alpha|$ happens. In our example the characteristic polynomial is given explicitly as
\begin{align}
D(z,r,\tau)= \frac{2N}{\tau}\int_{0}^{\infty} r' \exp{\left(-N\frac{r^2+r'^2}{\tau}\right)} I_{0}\left(\frac{2Nrr'}{\tau}\right)
\det\mathcal{M}(z,r')  \dd r' 
\end{align}
We consider the following scaling around the origin
\begin{align}
	r' = \theta N^{-7/6}, \qquad |z| = x N^{-1/6}, \\
	|\alpha| = x N^{-1/6}, \qquad \tau = t N^{-4/3} .
\end{align}
Inserting these formulas to the equation above we find
an asymptotic function for large $N$ and for  $|\alpha| = x N^{-1/6}$:
\begin{align}
D(z=x N^{-1/6}, r=0, \tau = t N^{-4/3}) \sim \frac{t}{2} + \frac{t\sqrt{\pi t}}{4x} \exp \left ( \frac{t}{4x^2} \right ) \text{erfc}\left ( -\frac{\sqrt{t}}{2x} \right ).
\end{align}
We see that it reveals a non-perturbative divergent character near $x=0$. For the normal
Ginibre case we do not find the second term and therefore we identify it as a consequence of the non-normal initial condition. 
Concluding this section, we note that the formula \eqref{nonherm} 
was obtained in \cite{FEINBERGZEE} for an initial matrix
\begin{align}
X_0^{C} = |\alpha| \diag (1, e^{2\pi i/N}, e^{4\pi i/N} ... ~ e^{2\pi  i (N-1)/N}),
\end{align}
with $N$ initial equidistant eigenvalues lying on a circle of radius $|\alpha|$. This result is not surprising since the matrix $X_0^{C}$ is unitarily equivalent to a circulant matrix
\begin{align}
(X_0^C)_{ij} \sim (X_0)_{ij} + \alpha \delta_{i,N} \delta_{j,1},
\end{align}
which in turn differs from the non-normal initial matrix considered in this paper by just one element whose effect can be neglected in the large $N$ limit.  We also mention that finite $N$ spectral densities and large deviations  have been recently studied in~\cite{MIREILLECHARLES}. 


\section{Spectral stochastic equations}
\label{stochasticeq}

An interesting issue arises, if the stochastic dynamics of the elements of  non-hermitian random matrices corresponds to some stochastic equation for the spectra of the underlying matrix. Such phenomenon holds in the case of Gaussian Unitary Ensembles (GUE), as well as in the case of Circular Unitary Ensembles (CUE), as shown in the seminal paper by Dyson~\cite{DYSON}. In the simpler case of GUE, the corresponding Langevin equation for real eigenvalues reads
\be
\dd\lambda_i =\dd B_i + \sum_{i \neq j} \frac{\dd t}{\lambda_i-\lambda_j},
\label{LangGUE}
\ee
where $\dd B_i$ reflects the Brownian dynamics and the second, "drift" term comes from the Van der Monde determinant.  We neglect the optional Ornstein-Uhlenbeck term $-\lambda_i \,\,\dd t$ on the r.h.s. of (\ref{LangGUE}) since it 
only contributes to freezing the diffusing end-points of the spectra in the stationary limit. The corresponding Smoluchowski-Fokker-Planck equation written for the resolvent $G(z,\tau)$ of the spectral density $\rho(\lambda, t)$, takes (after rescaling the time $\tau=Nt$  and in the large $N$ limit) the form of complex, inviscid Burgers equation, i.e. the eq. (\ref{BurgersGUE})~\cite{BN1,SPEICHERBIANE}.

Equations for non-hermitian Gaussian ensemble obtained in this work exhibit structural similarity to the hermitian "Burgulence", hence a question arises, if some stochastic dynamics for complex eigenvalues does exist as well.
Recently, Osada~\cite{OSADA1} had examined a stochastic two-dimensional Coulomb system defined as 
\be
\dd\lambda_i =\dd B^{(2)}_i + \sum_{i \neq j} \frac{\dd t}{\bar{\lambda}_i-\bar{\lambda}_j}=\dd B^{(2)}_i + \sum_{i \neq j} \frac{\lambda_i-\lambda_j}{|\lambda_i-\lambda_j|^2}\dd t,
\label{LangGIN}
\ee
where $\dd B^{(2)}_i$ is a two-dimensional (complex) Brownian walk. In particular, he has shown that such interacting Brownian motion equipped with trivial initial condition (all $\lambda_i$ put to zero at $t=0$) leads to limiting distribution  of $\lambda_i$ representing a uniform disk, therefore resembling the Ginibre Ensemble spectrum. 

From the point of view of our analysis this result is curious, since we have proven that the dynamics of eigenvalues is intimately connected to the dynamics of eigenvectors, which seem to decouple completely from ``hypothetical'' eigenvalues of Ginibre ensemble in (\ref{LangGIN}). We suggest the resolution of this puzzle. 

It is well known, that in so-called Normal Random Matrix model~\cite{CHAUZABORONSKI} with axially symmetric potentials all correlation functions can be expressed in terms of holomorphic functions of a single variable. Moreover, the exact integrability of such models can be linked to $(2+1)$-dimensional Burgers equations. Quite remarkably, in the case of the potential $V(z,\bar{z})=|z|^2$, the correlations of the Normal Random Matrix model are identical to the correlations of the Ginibre Ensemble~\cite{MEHTA}. We therefore conjecture, that the stochastic equation \eqref{LangGIN} corresponds rather to Gaussian Normal Random Matrix model than to the Ginibre Ensemble. 

Because normal matrices are diagonalizable by a single unitary transformation, alike the hermitian matrices, the eigenvectors decouple from the eigenvalues which explains the lack of these degrees of freedom in (\ref{LangGIN}) and its formal similarity to hermitian stochastic equation \eqref{LangGUE}. Based on this and the significance of eigenvectors presented in this paper we find it highly probable that the Ginibre-like dynamics of (\ref{LangGIN}) is accidental and proper stochastic equation governing the dynamics of non-hermitian random matrices is not known. 

The speculative links  between our approach for generic complex matrices and non-Gaussian Normal Random Matrix models represent a challenge, which we plan to address in the forthcoming publications. 

\section{Conclusions}
\label{conclusions}

We have shown that a consistent description of non-hermitian Gaussian ensemble requires the knowledge of the detailed dynamics of co-evolving eigenvalues and eigenvectors. Moreover, the dynamics of eigenvectors seems to play a superior role (at least in the $N \to \infty$ limit) and leads directly to the inference of the spectral properties. This is a dramatically different scenario as compared to the standard random matrix models, where the statistical properties of eigenvalues are of primary importance, and the properties of eigenvectors are basically trivial due to the their decoupling from the spectra.  We have considered examples of Ginibre, spiric section and non-normal Ginibre where the formulas for spectral density, the 1-point eigenvector correlation function, electrostatic potential and universal functions were obtained and positively crosschecked with numerical simulations. By studying the dynamics of characteristics we anticipated the novel universality obtained for the spiric case as an error function type. We obtained compact formulas for both spectral density and eigenvector correlator for which the latter unraveled a promising determinantal structure. The diffusion equation, as an equation exact for finite $N$, was used mainly to obtain the universal behavior.

We conjecture that the hidden dynamics of eigenvectors discovered by us and described for the Gausssian non-hermitian ensemble,
is a general feature of all non-hermitian random matrix models, and has to appear systematically in $1/N$ expansion.

Our formalism could  be exploited to expand the area of application of non-Hermitian random matrix ensembles within problems of growth, charged droplets in quantum Hall effect and gauge theory/geometry relations in string theory  beyond the subclass of complex matrices represented by normal matrices. 

One of the challenges is an explanation, why, despite being so different, the Smoluchowski-Fokker-Planck equations for hermitian and non-hermitian random matrix models exhibit structural similarity to simple models of turbulence, where so-called Burgers equation plays the vital role, establishing the flow of the spectral density of eigenvalues in the case of  the hermitian or unitary ensembles  and the flow of certain eigenvector correlator in the case of non-hermitian ensembles. 

We believe that our findings will contribute to understanding of several puzzles of non-hermitian dynamics, alike 
extreme sensitivity of spectra of non-hermitian systems to perturbations~\cite{CHALKERMEHLIG,FYODSAV}. We also hope that the  quaternion extension used in our paper may help to understand better the mathematical subtleties of the measure of non-hermitian operators~\cite{SNIADY}.

\section{Acknowledgments}
MAN, JG, ZB, WT and PW are supported by the Grant DEC-2011/02/A/ST1/00119 of the National Centre of Science. MAN thanks  Neil O'Connell for pointing  out the reference ~\cite{OSADA1} and appreciates discussions  with Neil O'Connell, Boris Khoruzhenko, Andu Nica, Dima Savin, Roger Tribe and Oleg Zaboronski. 
The authors  are also grateful to Charles Bordenave  for pointing  the relevant  reference~\cite{MIREILLECHARLES} on mathematical subtleties of normal Gaussian models.  JG thanks prof. Thomas Guhr for the warm hospitality at the University Duisburg-Essen where part of this work was done.


\appendix


\section{Derivation of the diffusion equation \eqref{ddiff}}
\label{appdiff}
In this appendix we demonstrate that the averaged extended characteristic polynomial 
(\ref{ddef}) obeys the diffusion equation (\ref{ddiff}). The determinant in (\ref{ddef})
can be expressed with help of Grassmann variables as
\begin{align}
	D(z,w,\tau) = \int \mathcal{D}[X] \mathcal{D}[\eta,\xi] P(X,\tau) \exp T_G(X,z,w;\eta,\xi),
\end{align}
with the object $T_G$ given by
\begin{align}
	T_G(X,z,w;\eta,\xi) = \sum_{i,j} \left ( - x_{ij} (\bar{\eta}_i \eta_j + \bar{\xi}_j \xi_i ) -i y_{ij} \left ( \bar{\eta}_i \eta_j - \bar{\xi}_j \xi_i \right ) \right ) + \sum_i \left ( z \bar{\eta}_i \eta_i + \bar{z} \bar{\xi}_i \xi_i + w \bar{\xi}_i \eta_i -\bar{w} \bar{\eta}_i \xi_i \right ),
\end{align}
where $\eta,\bar{\eta},\xi$ and $\bar{\xi}$ are Grassmann variables and 
$X_{ij} = x_{ij} + i y_{ij}$. With the help of heat equation \eqref{prob} for the joint probability density function $P(X,\tau)$ we obtain
\begin{align}
	\partial_\tau D(z,w,\tau) & = \int \mathcal{D} [X,\eta,\xi] \left (\partial_\tau P \right ) \exp T_G = \frac{1}{4N} \int \mathcal{D} [X,\eta,\xi] \left (\sum_{i,j} \left ( \partial^2_{x_{ij}} + \partial^2_{y_{ij}} \right ) P \right ) \exp T_G = \nonumber \\
	& = \frac{1}{4N} \int \mathcal{D} [X,\eta,\xi] P \left (\sum_{i,j} \left ( \partial^2_{x_{ij}} + \partial^2_{y_{ij}} \right ) \exp T_G \right ) = \frac{1}{N} \int \mathcal{D} [X,\eta,\xi] P \sum_{ij} \bar{\eta}_i \eta_j \bar{\xi}_j \xi_i \exp T_G ,
	\label{lhs}
\end{align}
where we integrated by parts twice and we used 
\begin{align}
	\partial^2_{x_{ij}} \exp T_G & = (\bar{\eta}_i \eta_j + \bar{\xi}_j \xi_i )(\bar{\eta}_i \eta_j + \bar{\xi}_j \xi_i ) \exp T_G = 2 \bar{\eta}_i \eta_j \bar{\xi}_j \xi_i \exp T_G, \\
	\partial^2_{y_{ij}} \exp T_G & = (\bar{\eta}_i \eta_j + \bar{\xi}_j \xi_i )(\bar{\eta}_i \eta_j + \bar{\xi}_j \xi_i ) \exp T_G = 2 \bar{\eta}_i \eta_j \bar{\xi}_j \xi_i \exp T_G .
\end{align}
On the other hand we have
\begin{align}
\label{rhs}
	\partial_{w\bar{w}} \exp T_G = \sum_{i,j} \bar{\eta}_i \eta_j \bar{\xi}_j \xi_i \exp T_G,
\end{align}
from which it follows that 
\begin{align}
\label{jos}
	\frac{1}{N} \partial_{w\bar{w}} D(z,w,\tau) =  \frac{1}{N} \int \mathcal{D} [X,\eta,\xi] P \sum_{ij} \bar{\eta}_i \eta_j \bar{\xi}_j \xi_i \exp T_G .
\end{align}
We see that the expressions on the right hand side of \eqref{lhs} and \eqref{jos} are identical and thus we have $\frac{1}{N} \partial_{w\bar{w}} D(z,w,\tau)=\partial_\tau D(z,w,\tau)$  \eqref{ddiff}. 
As a side remark we note that this calculation can be almost verbatim repeated 
for the averaged extended "inverse" characteristic polynomial
\begin{align}
	F(z,w,\tau) = \left < \det \left ( \begin{matrix} z - X & - \bar{w} \\
	 w & \bar{z} - X^\dagger  
	 \end{matrix} \right )^{-1} \right >_\tau,
\end{align}
which obeys the diffusion equation in the opposite time direction $\tau \rightarrow -\tau$
\begin{align}
	-\del_\tau F = \frac{1}{N} \del_{w\bar{w}} F.
\end{align}


\section{The $r \neq 0$ regime as a Wishart/chiral deformation}
\label{apprneq0}
The majority of results discussed in this paper are confined to the "physical" region where $r \to 0$. One can however keep the parameter $r$ nonzero. In the Coulomb gas interpretation, this deformation introduces a complex nonlinear interaction between the eigenvalues of unknown interpretation.

\begin{figure}[ht!]
  \centering
	\includegraphics[width=1.\textwidth]{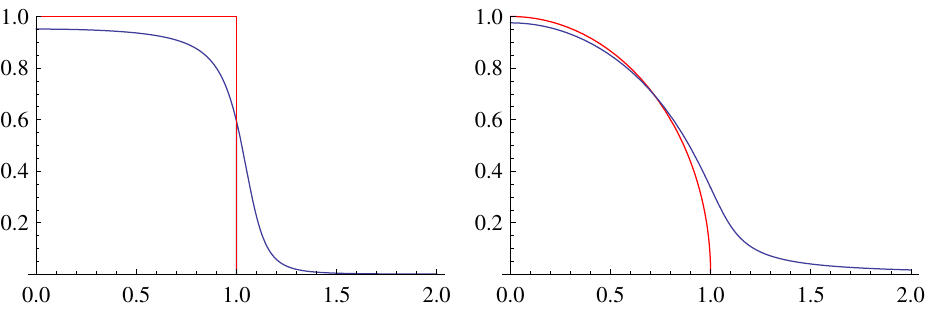}
      \caption{The large $N$ spectral density (left plot) and eigenvector correlator $|v_\tau|$ (right plot) for $r=0.05$ (blue line) and $r=0$ (red line).}
      \label{ggnzr}
\end{figure}

We consider the Ginibre case. The expression in the Hopf-Lax equation (\ref{hlr})
\begin{align}
	\phi(z,r,\tau) = \max_{r'} \left ( \ln (|z|^2 + r'^2) - \frac{(r'-r)^2}{\tau} \right ),
\end{align}
is maximized by $r_*$ obeying a cubic equation
\begin{align}
(r_*-r)(r_*^2+|z|^2) = r_* \tau,
\end{align}
The corresponding spectral density $\rho(z,r,\tau)$ and the 
eigenvector gradient $\nu(z,r,\tau)$ are shown in Fig. \ref{ggnzr}. There are no critical points for $r \ne 0$ and instead we see a smooth crossover between two phases. This is in accordance with the fact that the shock is present only on the $r=0$ plane.
Consider the formula for the averaged extended characteristic polynomial for the Ginibre
evolution 
\begin{align}
D(z,r,\tau)= \frac{2N}{\tau}\int_{0}^{\infty} r' \exp{\left(-N\frac{r^2+r'^2}{\tau}\right)} I_{0}\left(\frac{2Nrr'}{\tau}\right)
\left( |z|^2 + r'^2\right)^N\dd r' .
\end{align}
Applying the Newton binomial formula to $\left( |z|^2 + r'^2\right)^N$ 
an using the following integral representation of Laguerre polynomials
\begin{align}
	\int_0^\infty dq q^{2k+1} e^{-\frac{N}{\tau} q^2} I_0\left (\frac{2Nrq}{\tau} \right ) = \frac{\tau}{2N} \left ( \frac{\tau}{N} \right )^k k! L_k \left ( - \frac{N}{\tau} r^2 \right ) e^{\frac{N}{\tau} r^2}.
\end{align}
we get
\begin{align}
	D(z,r,\tau) = e^{-\frac{N}{\tau} r^2} N! \left ( \frac{\tau}{N} \right )^{N} \sum_{k=0}^N \frac{1}{k!} L_{N-k}\left ( - \frac{Nr^2}{\tau} \right ) \left ( \frac{N |z|^2}{\tau} \right )^k.
\end{align}
It is not surprising that Laguerre polynomials show up in this context since the the Ginibre ensemble is closely related to the Wishart/chiral ensemble where they naturally occur. In fact by recalling the definition \eqref{ddef}
\begin{align}
	D(z,r,\tau) = \left < r^2 + (z-X)(\bar{z}-X^\dagger) \right >_\tau
\end{align}
we see that for $z\to 0$ the object within the brackets is a Wishart matrix 
and
\begin{align}
	D(z=0,r,\tau) = e^{-\frac{N}{\tau} r^2} N! \left ( \frac{\tau}{N} \right )^{N} L_N\left (-\frac{N}{\tau} r^2 \right ) .
\end{align}
For $r\rightarrow 0$ we have  $D(z=0,r=0,\tau) = N! \left ( \frac{\tau}{N} \right )^{N}$.
\section{The kernel structure}
\label{appgg}

We argue that in the case of Ginibre matrices, the characteristic polynomial $D$ in the $r \to 0$ limit has essentially the same information as the microscopic kernel of the underlying determinantal process. This stems from the observation made in \cite{AKEMANN2} where the author connects the $n$-point correlation function of a Ginibre matrix model to a random matrix QCD partition function with $n$ quarks with appropriately decreased matrix size:
\begin{align}
	\det (K_N(z_i,z_j))_{i,j=1...n} = c(z_1,...,z_n) \left < \prod_{i=1}^n \det (z_i-X) \det (\bar{z}_i - X^\dagger ) \right >_{X_{N-n}},
\end{align}
where $\det (K_N(z_i,z_j))_{i,j=1...n} = \left < \prod_{i=1}^n \Tr \delta^2(z_i-X) \right >_{X_N} $ is the correlation function averaged over $N$ dimensional matrix and $c$ denotes a known $z$ dependent proportionality factor. 

We study closer the $n=1$ case because for this parameter the prefactor $c$  is the weight function $w(z)$ and the averaged determinants are exactly the characteristic polynomial $D$ in the $r \to 0$ limit. Therefore, in this particular case, we obtain a on-diagonal kernel formula
\begin{align}
	K_N(z,z) = C_N w(z) D^{(N-1)}(z,r=0,\tau),
\end{align}
with some numerical constant $C_N$. The off-diagonal kernel is not so easily obtainable by the above formula however we make an educated guess based on the symmetry of arguments and the result of Akemann and Vernizzi \cite{AKEMANN}. We write the full kernel as
\begin{align}
\label{akver}
	K_N(z,v) = C_N \sqrt{w(z)}\sqrt{w(v)} D^{(N-1)}([z,v],r=0,\tau),
\end{align}
where the two-argument characteristic polynomial $D$ was created by substituting $|z|^2 \to z \bar{v}$:
\be
\label{dnew}
D^{(N-1)}([z,v],r,\tau)= \frac{2N}{\tau} \int_{0}^{\infty}q \exp{\left(-N\frac{q^2+r^2}{\tau}\right)} I_{0}\left(\frac{2Nqr}{\tau}\right)(q^2+z\bar{v})^{N-1}  {\rm d}q, 
\ee
which for $r=0$ it is readily solved as
\begin{align}
\label{dtaunew}
	D^{(N-1)}([z,v],r=0,\tau) = 
	\left ( \frac{\tau}{N} \right )^{N-1} \Gamma(N) \sum_{k=0}^{N-1} \left (\frac{Nz\bar{v}}{\tau} \right )^k \frac{1}{k!},
\end{align}

To complete the argument demonstrating that the full information resides in the characteristic polynomial $D$ we will find an a priori unknown weight function $w(z)$ present in formula \eqref{akver} from the $D$ alone. This is done by using decomposition in the biorthogonal basis
\begin{align}
	D^{(N-1)} = \sum_{k,l=0}^{N-1} c_{kl} X_k(z) \bar{Y}_l(v),
\end{align}
where $X_i, Y_i$ are biorthogonal polynomials with respect to an unknown measure $W$
\begin{align}
\label{moments}
	\int d^2z W(z) X_i(z) \bar{Y}_j(z) = g_{ij}.
\end{align}
The expansion terms $c_{ij}$ and the bi-orthogonality matrix $g_{ij}$ are inverses $g_{ij} = c^{-1}_{ij}$. We  proceed by finding a basis $X_i, Y_j$ in which $c_{kl}$ are  diagonal. Then we infer the formula for $W$ by considering its moments \eqref{moments}. We start with a formula \eqref{dtaunew} which is already in a diagonal form with
\begin{align}
	X_k(z) & = z^k, \qquad Y_k(v) = v^k, \qquad c_{kk} = \frac{N!}{k!} \left ( \frac{\tau}{N} \right )^{N-1-k} .
\end{align}
The moments are therefore given by
\begin{align}
	\int d^2z~ W(z) |z|^{2k} & = \frac{k!}{N!} \left ( \frac{N}{\tau} \right )^{N-1-k}.
\end{align}
By assuming the radial symmetry and setting $|z|^2 = p$ we obtain
\begin{align}
	\int_0^\infty dp W(\sqrt{p}) p^k = \frac{(N/\tau)^N}{\pi N!} \frac{k!}{(N/\tau)^{k+1}}.
\end{align}
from which the characteristic function is given by
\begin{align}
	M_W(t) = \int_0^\infty dp W(\sqrt{p}) e^{itp} = \frac{(N/\tau)^{N-1}}{\pi N!} \left ( 1-\frac{it\tau}{N} \right )^{-1}.
\end{align}
It is exactly the characteristic function of an exponential distribution $\lambda e^{-\lambda x}$ with $\lambda=N/\tau$
\begin{align}
	W(\sqrt{p}) = \frac{(N/\tau)^{N}}{\pi N!} e^{-\frac{N}{\tau} p}.
\end{align}
This procedure gives the unknown weight $w(z) = e^{-N/\tau |z|^2}$ along with the normalization coefficient $C_N = \frac{1}{\pi N!} \left ( \frac{N}{\tau} \right )^{N}$
\begin{align}
	W(|z|^2) = C_N w(z).
\end{align}
The kernel is therefore equal to
\begin{align}
	K_N(z,v) = \frac{1}{\tau \pi} \exp \left ( - \frac{N}{2\tau} (|z|^2+|v|^2) \right ) \sum_{k=0}^{N-1} \left (\frac{Nz\bar{v}}{\tau} \right )^k \frac{1}{k!}.
\end{align}

\end{document}